\documentclass[a4paper, 11pt]
{article} 
\usepackage{tikz}

\usepackage{afterpage}
\usepackage[utf8]{inputenc}
\usepackage[margin=1in]{geometry} 
\title{Anomaly Detection on Financial Time Series by Principal Component Analysis and Neural Networks\footnote{Python notebooks reproducing the results of this paper are available  
on \url{https://github.com/MadharNisrine/PCANN}. The authors would like to thank Pascal Oswald, Leader Market \& Counterparty Risks Modelling, from Natixis, for insightful discussions.}}
\usepackage[english]{babel}
\usepackage{dsfont}

\author{
\href{https://perso.lpsm.paris/~crepey/}{S. Cr\'epey}\thanks{{\tt stephane.crepey@lpsm.paris}. LPSM, Universit\'e Paris Cit\'e, 
France.} , 
N. Lehdili\thanks{{\tt noureddine.lehdili@natixis.com}. Expert Leader Market \& Counterparty Risks Modelling.} , 
N. Madhar\thanks{Email: {\tt nisrine.madhar@lpsm.paris}. PhD student, Universit\'e Paris Cité, 
France. The research of N. Madhar is funded by a CIFRE grant from Natixis.} , 
M. Thomas\thanks{{\tt maud.thomas@sorbonne-universite.fr}. CNRS, Laboratoire de Probabilités, Statistique et Modélisation, LPSM, Sorbonne Université, France.}  
}
\date{\today}

\usepackage{natbib}
\usepackage[colorlinks=false,linkcolor=black, citecolor=black, urlcolor=black]{hyperref}
\usepackage{siunitx}
\usepackage{amsfonts}

\usepackage{multicol}
\usepackage{url}            
\usepackage{booktabs}       
\usepackage[toc]{appendix}
\usepackage{subfig}
\usepackage{tikz}
\usetikzlibrary{positioning}
\usetikzlibrary{shapes,arrows}

\usepackage{amssymb}
\usepackage{amsfonts}
\usepackage{amssymb}
\usepackage{bbm}
\usepackage{algorithm}

\usepackage{ragged2e}
\usepackage{nicefrac}       
\usepackage{microtype}      
\usepackage[T1]{fontenc}    
\usepackage{stmaryrd} 
\usepackage{bm} 
\usepackage{makecell} 
\usepackage{caption} 
\usepackage{graphicx} 
\usepackage{verbatim} 
\usepackage[algo2e]{algorithm2e} 

\usepackage{mathtools}  
\usepackage{amssymb}  
\usepackage{xcolor}

\usepackage{comment}

\newcommand \blue[1]{\textcolor[rgb]{0,0,1}{#1}}
\def\shortciten#1{\blue{\shortciten{#1}}}

\newtheorem{assumption}{Assumption}

\hypersetup{
 colorlinks  = true, 
 urlcolor   = black, 
 linkcolor  = black, 
 citecolor  = black 
}
\makeatletter
\renewcommand\@makefnmark{\hbox{\@textsuperscript{\normalfont\color{blue}\@thefnmark}}}
\renewcommand\@makefntext[1]{%
 \parindent 1em\noindent
      \hb@xt@1.8em{%
        \hss\@textsuperscript{\normalfont\@thefnmark}}#1}
\makeatother

\newcommand{\Table}[1]{\text{Table #1}}

\begin{document}
\maketitle

\begin{abstract}

A major concern when dealing with financial time series involving a wide variety of
market risk factors is the presence of anomalies. These induce a miscalibration of the models used
to quantify and manage risk, resulting in potential erroneous risk measures. We propose an approach that aims at improving anomaly detection on financial time
series, overcoming most of the inherent difficulties. Valuable
features are extracted from the time series by 
compressing and reconstructing the data through principal component analysis.
We then define an anomaly score using a feedforward neural network. A time series is considered to be
contaminated when its anomaly score exceeds a given cut-off value. This cutoff value is not
a hand-set parameter, but rather calibrated as a neural network parameter, throughout the
minimization of a customized loss function. The efficiency of the proposed approach compared
to several well-known anomaly detection algorithms is numerically demonstrated on both synthetic and real data sets, with high and stable performance being achieved with the PCA NN approach. We show that value-at-risk estimation errors are reduced when
the proposed anomaly detection model is used with a basic imputation approach
to correct the anomaly.
   \medskip
   

\end{abstract}



\section{Introduction}

 In the context of financial risk management, financial risk models are of utmost importance in order to quantify and manage financial risk. Their outputs, risk measurements, can both help in the process of decision making and ensure that the regulatory requirements are met \citep{BISFRTB}. Financial management thus heavily relies on financial risk models and the interpretation of their outputs. The data usually consist of time series
representing a wide variety of market risk factors. A major issue of such data is the presence of anomalies. A value of the time series is considered to be abnormal whenever its behaviour is significantly different from the behaviour of the rest of the time series \citep{hawkins1980identification}. In this work, we focus on the detection of abnormal observations in a market risk factor time series used to calibrate financial risk models. Indeed, erroneous input data may wrongly impact the calibrated model parameters. A typical example is the estimation of the covariance matrix of a bank market risk factors. This covariance matrix is involved in the computation of value-at risks (VaR) or expected shortfalls (expected losses above the VaR level). Since the true covariance is unknown, it has to be estimated from the data. However, the presence of anomalies in the data might have an impact on this estimation. For this specific case, robust methods, less sensitive to anomalies, can be used. Yet, existing robust estimators are computationally expensive, with a polynomial or even exponential time complexity in terms of the number of market risk factors. A faster approach was suggested by \citet{cheng2019faster}, but this algorithm only applies to the estimation of the covariance matrix in the case of Gaussian distribution. Financial risk models used by banks are widespread and various. Therefore, instead of seeking for a robust version for each of them, we propose to detect anomalies directly on the time series. 

\paragraph{Contributions of the Paper}
Our PCA NN methodology identifies anomalies on time series using a principal component analysis (PCA) and neural networks (NN) as underlying models. This methodology overcomes common pitfalls associated with anomaly detection. PCA is first used as a features extractor on (augmented if needed) data. Anomaly detection is then performed in two steps. The first step identifies the time series with anomalies evaluating the propensity of the time series to be contaminated, as reflected by its so-called anomaly score. Toward this end, we calibrate a feedforward neural network through the minimization of a customized loss. 
This customized loss allows us to calibrate the cut-off value on the anomaly scores, without resorting to expert judgement. In this way, we remove the expert bias. The second step localizes the anomaly among the observed values of the identified contaminated time series.

\paragraph{Outline}
The PCA NN approach is detailed in Section \ref{section: model}. Section \ref{section: data} describes the methodology used for generating the data on which our approach is thoroughly tested in Sections 
\ref{section: model_eval}, \ref{section: model_eval2} and
\ref{section: model_eval3}, exploiting the knowledge of the data generating process for benchmarking purposes.
Section \ref{section:VaR} illustrates the benefit of using our approach as a data cleaning preprocessing stage on a downstream task, namely  value-at-risk computations. 
This is completed by further numerical experiments on real data sets in Section \ref{s:real}.
Section \ref{section: conclusion} concludes.
Section \ref{literatureReview} and \ref{apx:AppendixA} provide reviews of the anomaly detection literature and algorithms. 
Section \ref{apx:AppendixB} addresses the data stationarity issue.


%

\section{The PCA NN Anomaly Detection Approach}\label{section: model}


\subsection{Notations}\label{subsec:MethodNotation}
Let 

\begin{equation}\label{def:X}
\bm{X} = 
\left(
X^{1}, X^{2}, \ldots , X^{i},  \ldots, X^{n}
\right)^\top,
\end{equation}
where the column-vector $X^i= \left(x_{t_1}^i,x_{t_2}^i, \ldots, x_{t_j}^i\ldots, x_{t_{p}}^i\right) \in \mathbb{R}^{p}$ corresponds to the $i$-th observed time series, and $x_{t_j}^i$ to the value observed at time $t_j$ in the $i$-th time series. 




Our method fits into the supervised framework. We thus assume that the data matrix $\bm{X}$ comes along with two label vectors. The first label vector $\bm{A} = \left(A^1,\ldots,A^n\right) \in \{0,1\}^n$ identifies the time series containing anomalies referred to as contaminated time series. For $i=1,\ldots, n$, we define the identification labels as
\begin{align}
A^{i} = \begin{cases}
       1  & \text{if there exists $j$ such that $x^i_{t_j}$ is an anomaly}. \\
     0 & \text{otherwise.}
\end{cases}  
\label{eq:identLabel}
\end{align}

The second label vector $\bm{L}$ concerns solely the contaminated time series. Its coefficients correspond to the localization labels, i.e. time stamps at which an anomaly occurs. For $j=1,\ldots, p$ and $i \in I_c = \{i : A^i = 1\}$ (the set of contaminated time series),
\[
L^{i} =  \jmath  \quad \text{ when the anomaly occurs at time $t_{\jmath}$ for the $i$-th contaminated time series}, 
\]

$\bm{X},\bm{A}$ and $\bm{L}$ are general notations. As the model involves a learning phase, we denote by $\bm{X}^{Train}$, $\bm{A}^{Train}$ and $\bm{L}^{Train}$ the data used for calibration and, by $\bm{X}^{Test}$, $\bm{A}^{Test}$ and $\bm{L}^{Test}$ the independent data sets on which the model performance is evaluated.

\subsection{Methodology Overview} 
\label{subsection:modeldescription}
 
The anomaly detection model we propose is a two step supervised-learning approach. The contaminated time series identification step aims at identifying among the observed time series the ones with potential anomalies. The anomaly localization step  consists in finding the location of the anomaly in each contaminated time series identified during the first step. 

The model used in the first step falls under the scope of binary classification models, assigning to each time series a predicted label
\[
\widehat{A}^{i} = \begin{cases}
       1  & \text{if $X^{i}$ is considered as contaminated by the identification model.} \\
     0 & \text{otherwise.}
\end{cases}
\]

The second step uses a multi-classification model. For each contaminated time series identified in the first step, the model predicts a unique time stamp $\widehat{L}^{i}$ at which the anomaly has occurred. Formally, the observed value at the time stamp $t_{\hat{\jmath}^{i}}$ of the $i$-th contaminated time series is considered abnormal, meaning that the observation $x^i_{t_{\hat{\jmath}^i}}$ is abnormal. Thus, 

\[
\widehat{L}^{i} =  \hat{\jmath}^{i} \quad   \text{ with } \quad \hat{\jmath}^{i} \in \{ 1,\dots,p \}.
\]

Our two-step anomaly detection model localizes only one anomaly per run, if any. Several iterations allow removing all anomalies. Indeed, as long as the time series is identified as contaminated, by the first step of the method, the time series can go through the second step. Once all anomalies have been localized the final stage of the approach is to remove the abnormal values and suggest imputation values (see Section \ref{subsection:imputation}).

\subsection{Theoretical Basics of Principal Component Analysis}
\label{subsec:PCA}
A fundamental intuition behind our anomaly detection algorithm is the existence of a lower dimensional subspace where normal and abnormal observations are easily distinguishable. Since the selected principal components explain a given level of the variance of the data, they represent the common and essential characteristics to all observations. As one might reasonably expect, these characteristics mostly represent the normal observations. When the inverse transformation is applied, the model successfully reconstructs the normal observations with the help of the extracted characteristics, while it fails into reconstructing the anomalies.

In this section, we recall the theoretical tenets of the principal component analysis (PCA) useful for our purpose. The idea of PCA is to project the data from the observation space of dimension $p$ into a latent space of dimension $k$, with $k <p$. This latent space is generated by the $k$ directions for which the variance retained under projection is maximal. 

The first principal component $U$ is given by $U = w^\top\bm{X},$
where $w$ is a to-be-determined vector of weights. Since the PCA aims at maximizing the variance along its principal components, $w$ is sought for as 
\[
\arg \max_{\| w\| = 1}  w^\top\Sigma w \,,  
\]where $\Sigma$ is the covariance matrix of $\bm{X}$.
The optimal solution by the Lagrangian method is given by 
$\Sigma w = \lambda w \,$, where $\lambda>0$ is a Lagrangian multiplier. Multiplying both sides by $w^\top$ results in $w^\top\Sigma w = \lambda$. Following the same process with an additional constraint regarding the orthogonality between the principal components, one can show that the $k$ first principal components correspond to the eigenvectors associated with the $k$ dominant eigenvalues of $\Sigma$.






The transfer matrix $\Omega_{\bm{X}} \in \mathbb{R}^{k\times p}$ is defined by the eigenvectors of $\Sigma$ associated with its $k$ largest eigenvalues. This transfer matrix $\Omega_{\bm{X}}$ applied to any observation in the original observation space projects it into the latent space. Since the transformation is linear, reconstructing the initial observations from their equivalent in the latent space is straightforward. Using PCA on the data matrix $\bm{X}$, the transfer matrix $\Omega_{\bm{X}}$ can be inferred. The projection of the observations into the latent space $\bm{Z} \in \mathbb{R}^{n\times k}$ is then given by $\bm{Z} = \bm{X} \Omega_{\bm{X}}^\top $, while their reconstructed values are given by $\widehat{\bm{X}} = \bm{Z}\Omega_{\bm{X}}$.

The reconstruction errors $\varepsilon^i$ for each time series defined by 
\begin{align}
    \varepsilon^{i} =\widehat{X}^{i} - X^{i}, \quad i=1,\ldots, n, 
    \label{eq:recerror}
\end{align}
referred in the sequel as the new features, are our new representation of data, given as inputs to our models. Algorithm \ref{algo: FeaturesExtraction} summarizes how to compute these reconstruction errors. The two steps of the model use the same inputs, namely these reconstruction errors. Nevertheless, the processing these inputs undergo in each step differs.

     
\begin{algorithm}[H]
     \SetAlgoLined
     \caption{Features extraction with PCA }
     \label{algo: FeaturesExtraction}
     \SetKwInOut{Require}{inputs}
     \Require{number of principal components $k$\\out of sample data $\bm{X}^{Test}$\\training data if not trained yet $\bm{X}^{Train}$ }
     
\If{not trained yet}{
             $\Omega_{\bm{X}^{Train}} \leftarrow \mathrm{PCA}(\bm{X}^{Train},k)$\;
         }
     
    $\bm{\varepsilon}^{Train} = \bm{X}^{Train} \left(\Omega_{\bm{X}^{Train}}^\top\Omega_{\bm{X}^{Train}} - I_{p}\right)$
    
    $\bm{\varepsilon}^{Test} = \bm{X}^{Test} \left(\Omega_{\bm{X}^{Train}}^\top\Omega_{\bm{X}^{Train}} - I_{p}\right)$\\
     
     \SetKwInOut{Output}{outputs}
     \Output{$\bm{\varepsilon}^{Train}$ reconstruction errors for the train set \\
     $\bm{\varepsilon}^{Test}$ reconstruction errors for the test  set} 
\end{algorithm}
     




\subsection{Contaminated Time Series Identification}\label{subsec:Step1}

The first step of the approach aims at identifying the time series with anomalies based on the reconstruction errors obtained using PCA and Algorithm \ref{algo: FeaturesExtraction}. In this section, we detail the process these new features undergo. We assign to each reconstruction error an anomaly score which computation is handled by a neural network (NN) \citep[see][]{goodfellow2016deep}. We also numerically demonstrate that our NN has an advantage over a naive approach in terms of being able to assess more accurately the propensity of a time series of being contaminated. 


\paragraph{Naive approach}

A naive and natural approach to define the anomaly score is to consider the $\ell^2$-norm of the reconstruction errors $\widetilde{\varepsilon}_i = \lVert \varepsilon^{i}\rVert_2$, $i=1,\ldots, n$. This anomaly score represents the propensity of a time series to be contaminated or not.
The scores are first split into two sets depending on whether the corresponding time series is contaminated or not. We denote by $\widetilde{\bm{\varepsilon}}^c$ and $\widetilde{\bm{\varepsilon}}^u$ the set of contaminated and uncontaminated time series anomaly scores, i.e.


\begin{align*}
    \widetilde{\bm{\varepsilon}}^c &:= \{ \lVert \varepsilon^{i}\rVert_2; \;A^{i} = 1\,  \} = \{ \widetilde{\varepsilon}_{i}, \,   i\in  I_c \} \\
    \widetilde{\bm{\varepsilon}}^u &:= \{ \lVert \varepsilon^{i}\rVert_2; \;  A^{i} = 0\, \} = \{ \widetilde{\varepsilon}_{i}, \,   i\notin I_c \} .
\end{align*}

The density distribution function of each class of time series $f^u$ and $f^c$, i.e. for both $\widetilde{\bm{\varepsilon}}^c$ and $\widetilde{\bm{\varepsilon}}^u$, is then estimated using kernel density estimation \citep{wkeglarczyk2018kernel}. For $l=\{c,u\}$, let $\bm{\varepsilon}^l$  be an i.i.d sample of $n^l$ observations from a population with unknown density $f^l$. The corresponding kernel estimator is given by

\begin{align}
    \hat{f}^l(s) = \frac{1}{n^l\mathfrak{h}} \sum_{k=1}^{n^l} \mathcal{K}\left( \frac{s - \widetilde{\varepsilon}_k^l}{\mathfrak{h}} \right),
    \label{kde}
\end{align}
 where $\mathcal{K}$ is a kernel function and $\mathfrak{h}$ is a smoothing parameter.

The cut-off value $s$ is then chosen as the intersection between the two empirical density functions, i.e.

\[
    \hat s= \arg\min_s\left\{\lvert \hat{f}^u(s) - \hat{f}^c(s) \rvert <  \eta \right\} \, ,
\]
where $\eta$ is a to be tuned precision level. The selected cut-off value $\hat s$ represents the value of the score for which the area under the curve of the density of uncontaminated time series above $\hat s$ and the area under the curve of contaminated time series below $\hat s$ are as small as possible. We expect it to correspond to a value exceeded by a relative low number of $\widetilde{\varepsilon}_i$ with $i \notin I_c$ as $\hat{f}^u(s)$ is expected to be flat around the cut-off value exceedance region, and exceeded only by few $\widetilde{\varepsilon}_i$ with $i \in I_c$.  

\begin{figure}
 \centering{
  \begin{tabular}{cc}
  \includegraphics[width=0.45\linewidth]{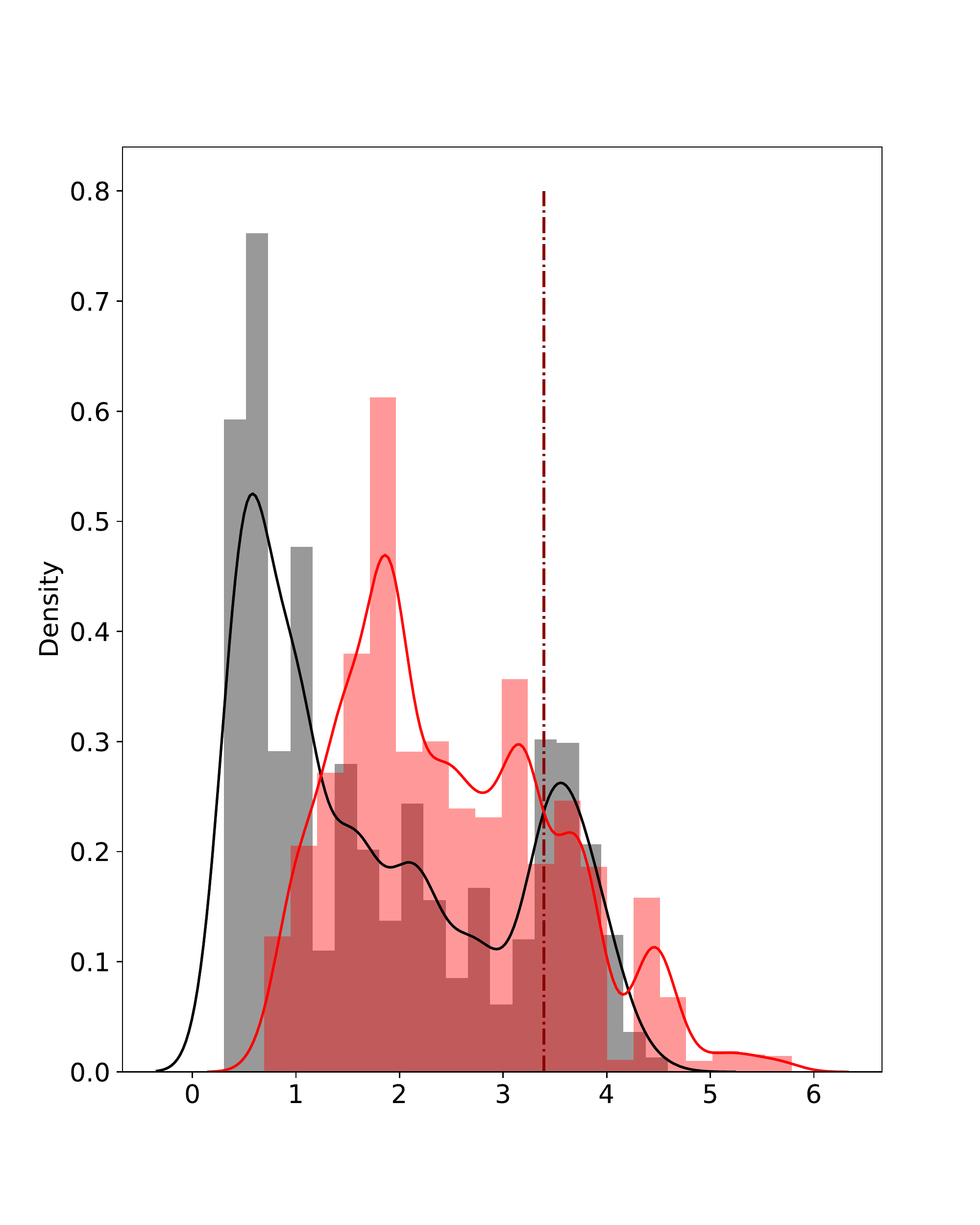} &
  \includegraphics[width=0.45\textwidth]{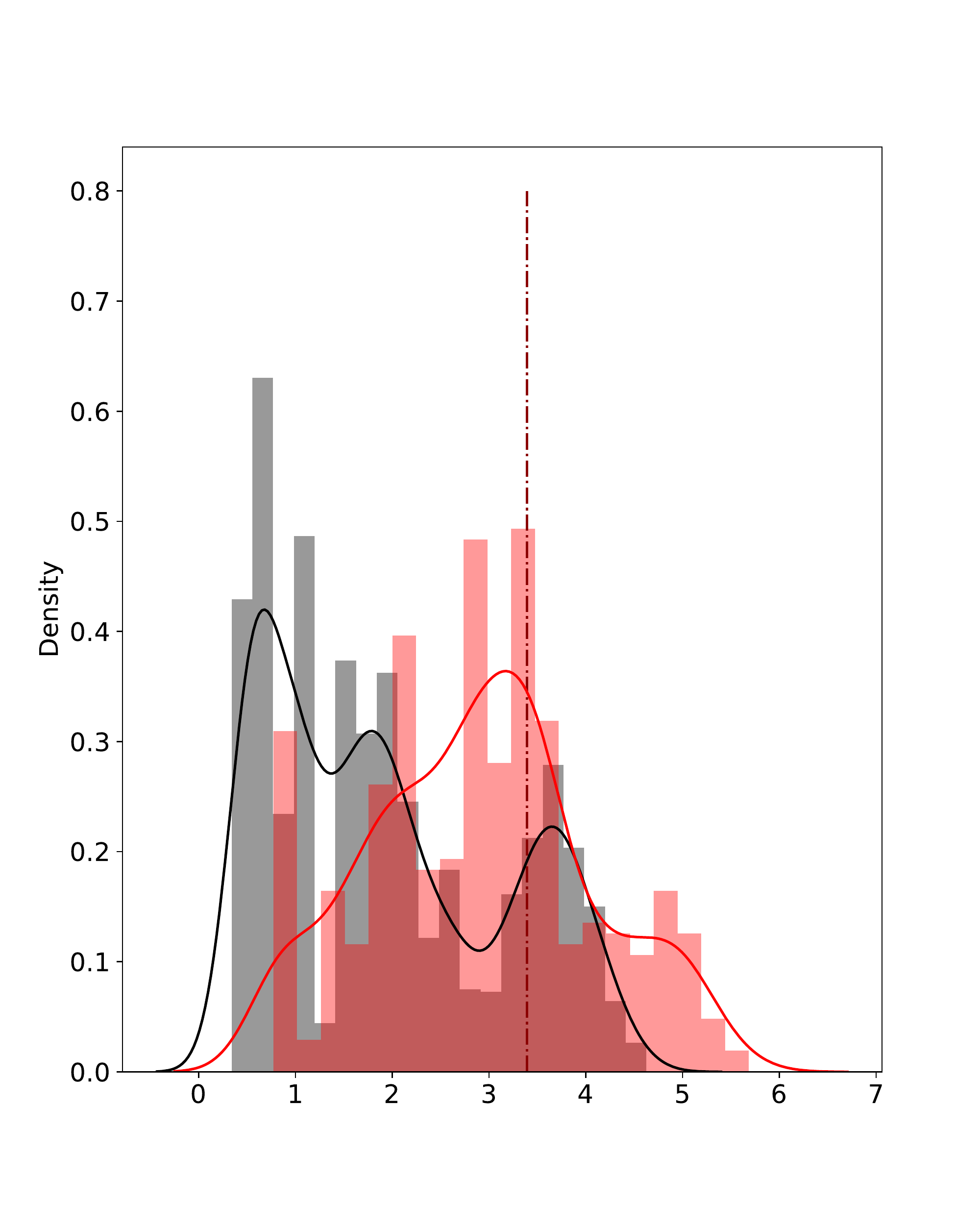}
  \end{tabular}}
  \caption{Empirical densities of anomaly scores given by the naive approach for uncontaminated time series in black and contaminated time series in red,  on the train set \textit{(left)} and  on a test set \textit{(right)}. The dotted dark red line represents the cut-off value.} 
  \label{fig:DensityPCAReconsErrorNaive}
\end{figure}

As shown in Figure \ref{fig:DensityPCAReconsErrorNaive}, the naive approach results in a non-negligible overlapping region between the two densities, both on the train set (left) and for the test set (right). This region represents the anomaly score associated with time series that could be either contaminated or uncontaminated. The uncertainty regarding the nature of the observation when its anomaly score belongs to this region is relatively high in comparison with observations whose anomaly scores lie on the extreme left-hand or right-hand side of the calibrated cut-off value. Moreover, because we are not able to provide a clear separation between the scores of uncontaminated and contaminated time series, we may expect the model to mislabel future observations, resulting in a high rate of false positives and true negatives.

\paragraph{Neural network approach}

In view of getting a clearer separation of the densities, we propose an alternative approach for the computation of the anomaly scores built upon the reconstruction errors. Let $F$ denote the function that associates with each reconstruction error $\varepsilon^{i}$ its anomaly score. In the naive approach, $F$ corresponds to the $\ell^2$-norm. Hereafter, $F$ instead represents the outputs of a feedforward Neural Network (NN). The training of the corresponding NN aims at minimizing a loss function that reflects our ambition to construct a function $F$ giving accurate anomaly scores. The network used for computing the anomaly scores is defined by
\begin{align*}
    &F(\varepsilon) = \left (h^H \circ h^{H-1} \circ \dots h^2 \circ h^1\right )(\varepsilon),
\end{align*}
where $h^i(\varepsilon) =\left(W^i \cdot  \varepsilon + b^i \right)^+$
 represents the computation carried in the $i$-th layer of the neural network with $n_h^i$ hidden units. $H$ stands for the number of layers of the NN, with ReLU activation applied element-wise. We finally estimate the labels from the outputs following $\bm{\hat{A}} = \mathds{1}_{ \left\{(F(\bm{\varepsilon}) - s)^+ > 0\right\}}$.

The idea of the learning is to calibrate the weights $\bm{W} := \left \{W^i \in \mathbb{R}^{n_h^{i} \times n_h^{i-1}  }, i =2, \dots, H \right \}$ and the biases $\bm{b}:= \left \{b^i \in \mathbb{R}^{n_h^{i}}, i= 1, \dots, H\right \}$ such that the NN is able to accurately assess to which extent a time series is contaminated, with a clear distinction between the anomaly scores assigned to uncontaminated time series and those assigned to the contaminated ones, while integrating the calibration of the cut-off value $s$ as part of the learning. We denote by $\Theta:=\{\bm{W},\bm{b},s\}$ the set of parameters to be calibrated through the NN training.  To meet these needs, the loss we minimize during the learning is given by 

\begin{equation}
 \mathcal{L}_{\Theta}\left(\bm{A}, \hat{\bm{A}}\right) = \mathrm{BCE}(\bm{A},\bm{\hat{A}}) + \mathrm{AUC Density}^u_{\bm{A},F(\bm{\varepsilon})} + \mathrm{AUC Density}^c_{\bm{A},F(\bm{\varepsilon})},
    \label{eq:myLoss}
\end{equation}
 where 
\begin{align*}
     BCE(\bm{A},\bm{\hat{A}})  = - \frac{1}{n}\sum_{i}\left(A^{i} \log(\hat{A}^{i}) + (1-A^{i}) \log(1-\hat{A}^{i})\right) \, 
\end{align*}
is the binary cross-entropy, a well-known loss function classically used for classification problems. To this first component of our loss function, we add two components that aim at downsizing the overlapping region between the density of anomaly score of both types of observations. In order to have a control on the latter, we consider $\mathrm{AUC Density}^u$ and $\mathrm{AUC Dentsity}^c$, which correspond to the area under the curve of the probability density function of anomaly scores the model assigns to contaminated and uncontaminated observations, respectively,\textbf{}  i.e
\begin{align}
    & \mathrm{AUC Density}^u(s) = \int_s^{\infty} \hat{f}^u_{\bm{A},F(\bm{\varepsilon})}(\omega) d\omega \; ,  \quad \quad  \quad  & \mathrm{AUC Density}^c(s) = \int_{-\infty}^s \hat{f}^c_{\bm{A},F(\bm{\varepsilon})}(\omega) d\omega \,. 
    \label{eq:AUCs}
\end{align}
The bounds of these integrals depend on the cut-off value $s$ and define the region for which we want the probability density functions to be as small as possible, which allows us to estimate $\hat{s}$. The estimated probability density functions $\hat{f}^u_{A,F(\bm{\varepsilon})}$ and $\hat{f}^c_{A,F(\bm{\varepsilon})}$  depend on $F(\bm{\varepsilon})$, the scores assigned to contaminated time series identification model, and of the identification labels $\bm{A}$. We describe with Algorithm \ref{algo: identificationLearning}, the scoring and cut-off value calibration achieved by the calibration of a feedforward network with the loss \eqref{eq:myLoss}. Each update of $\Theta$, AdamStep in Algorithm \ref{algo: identificationLearning}, is carried following the Adam optimization algorithm of \citep{kingma2014adam}.

\begin{algorithm}[H]
     \SetAlgoLined
     \caption{Scoring and cut-off calibration }
     \label{algo: identificationLearning}

     \SetKwInOut{Require}{inputs}
     \Require{learning rate $lr$\\
     number of maximum iterations $K$\\
     kernel density estimator parameters $\mathcal{K}$,$\mathfrak{h}$\\
     training data $\bm{\varepsilon}^{Train},\bm{A}^{Train}$\\}
     
    Initialize parameter $\Theta$, $\widehat{\mathcal{L}} = \infty$ and count index $k = 0$\;

     \While(){$k< K$}{

        $\\ scores \leftarrow F_{\Theta} \left(\bm{\varepsilon}^{Train}\right)$ \;\\
        $scores^{c} \leftarrow scores \mathds{1}_{\bm{A}^{Train}=1}$\;\\
        $scores^{u} \leftarrow scores \mathds{1}_{\bm{A}^{Train}=0}$\;  \tcp*[f]{Density estimation following \eqref{kde}} \; \\
        $\hat{f}^u_{\bm{A}^{Train},scores^u} = \mathrm{KernelDensityEstimator}(\mathcal{K},\mathfrak{h},scores^u)$ \; \\
      $\hat{f}^c_{\bm{A}^{Train},scores^c} = \mathrm{KernelDensityEstimator}(\mathcal{K},\mathfrak{h},scores^c)$\\
       \tcp*[f]{Loss evaluation following \eqref{eq:myLoss} and \eqref{eq:AUCs}} \; \\
      $\mathrm{AUCDensity}^u \leftarrow \mathrm{NumericalIntegration}(\hat{f}^u_{\bm{A}^{Train},scores^u},s)$ \\
      $\mathrm{AUCDensity}^c \leftarrow \mathrm{NumericalIntegration}(\hat{f}^c_{\bm{A}^{Train},scores^c},s)$\\
      $\bm{\hat{A}} \leftarrow \mathds{1}_{scores>s}$\\
      $\mathcal{L} \leftarrow \mathrm{BCE}\left(\bm{A}^{Train},\bm{\hat{A}}\right) +\mathrm{AUCDensity}^u + \mathrm{AUCDensity}^c  $\\
         \If(){ $\widehat{\mathcal{L}} > \mathcal{L}$ }{
             $\\\widehat{\mathcal{L}} \leftarrow \mathcal{L}$\; \\
             $\widehat{\Theta} \leftarrow \Theta$\; 
         }
         $\\ \Theta \leftarrow \mathrm{AdamStep}(\mathcal{L},\Theta,lr)$\; \\
         $k \leftarrow k +1$\;
     }
     \SetKwInOut{Output}{outputs}
     \Output{best calibrated parameter $\widehat{\Theta}=\{\widehat{\bm{W}},\hat{\bm{b}}, \hat{s}\}$.} \end{algorithm}

\begin{figure}
 \centering{
  \begin{tabular}{cc}
  \includegraphics[width=0.45\linewidth]{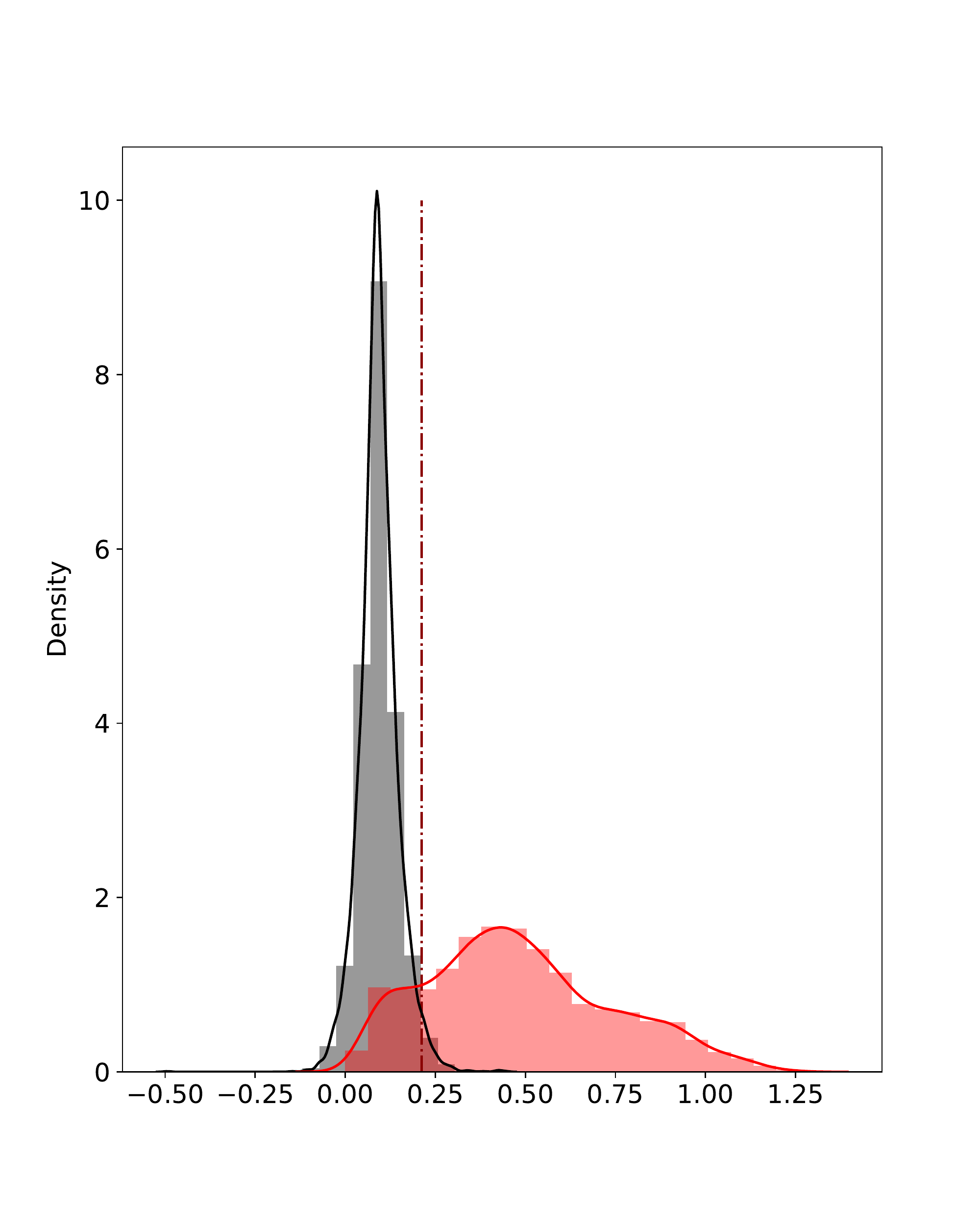} &
  \includegraphics[width=0.45\textwidth]{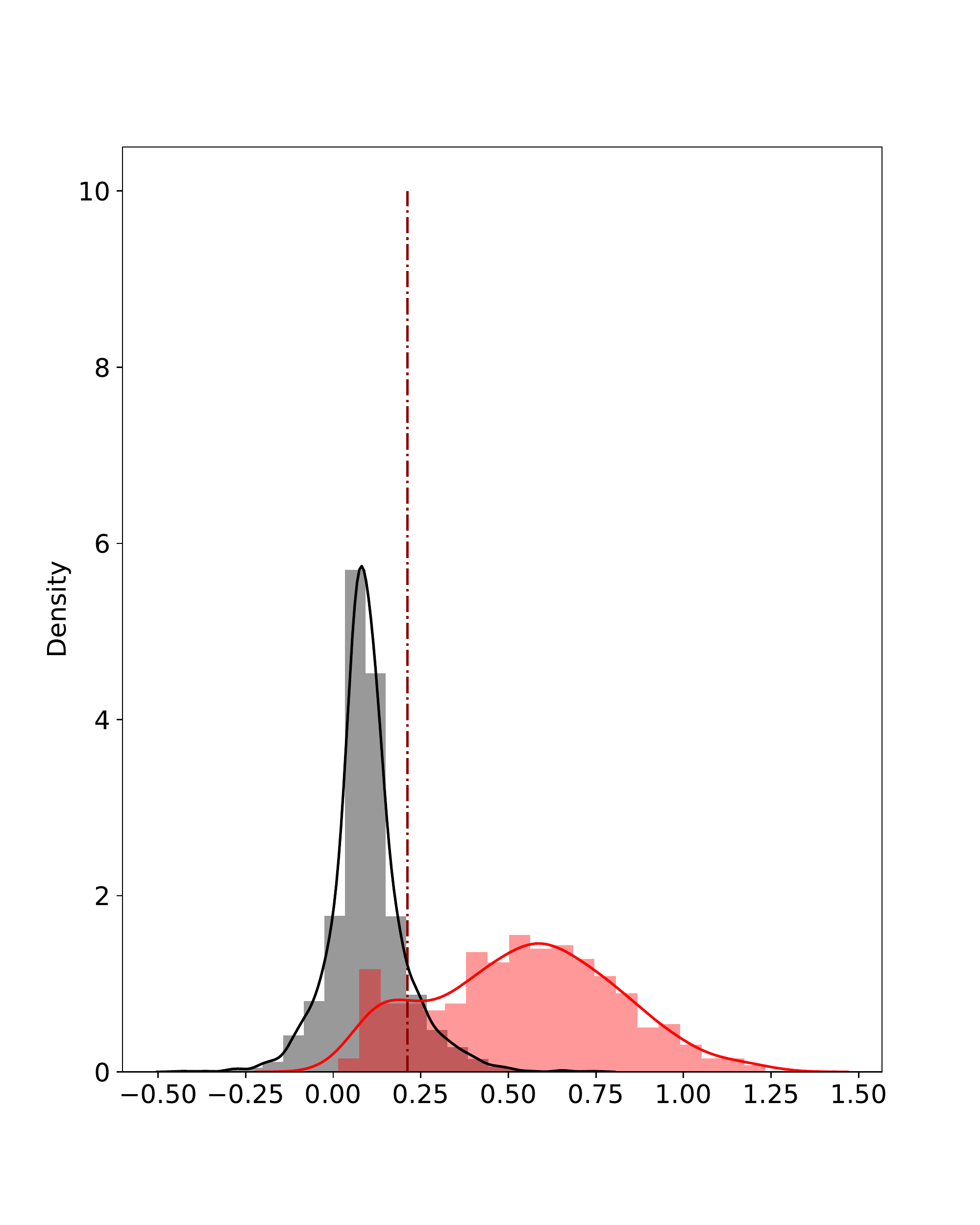}
  \end{tabular}}
  \caption{Empirical densities of anomaly scores given by the NN approach for uncontaminated time series in black and contaminated time series in red, on the train set \textit{(left)} and on a test set \textit{(right)}. The dotted dark red line represents the calibrated cut-off value $\hat{s}$.} 
  \label{fig:DensityPCAReconsErrorNN}
\end{figure}

\begin{table}[H]
\centering
\begin{tabular}{c}

\begin{tabular}{c|c|c}
    Approach    & $\mathrm{AUC Density}^u$ & $\mathrm{AUC Density}^c$\\
    \hline
Naive & 0.1725    & 0.7898     \\ 
\hline
NN    &0.05153    & 0.1550   
\end{tabular}
\begin{tabular}{c|c|c}
    Approach        & $\mathrm{AUC Density}^u$ & $\mathrm{AUC Density}^c$\\ 
    \hline
Naive & 0.1897    & 0.6354   \\ \hline
NN    & 0.1290    & 0.1367    
\end{tabular}

\end{tabular}
\caption{AUC obtained with the naive and the NN approaches for the train set \textit{(left)} and the test set \textit{(right)}. }
\label{tab:AUC}
\end{table}

Table \ref{tab:AUC} shows that with the NN approach the densities of scores assigned to each type of time series display the expected behaviours on the left-hand (right-hand) side of the cut-off value for contaminated (uncontaminated) time series, on both the train and test sets. Actually, the lower $\mathrm{AUC Density}^u$ ($\mathrm{AUC Density}^c$), the lower the number of uncontaminated (contaminated) time series to which are assigned anomaly scores above (below) the cut-off value,  which prevents mislabelling.


Once the features $\bm{\varepsilon}$ and the calibrated NN are provided, the contaminated times series identification model is ready for use. This step is described by Algorithm \ref{algo: IdentificationModel}. 
  \begin{algorithm}[H]
     
     \SetAlgoLined
     \caption{Contaminated time series identification model }
    \label{algo: IdentificationModel}
     \SetKwInOut{Require}{inputs}
     \Require{time series to analyze $X^i$,\\ calibrated model parameter $\widehat{\Theta}$, \\}
    $\varepsilon^{i} \leftarrow \mathrm{PCAFeaturesExtraction}(X^i)$\;  \tcp*{cf. Algorithm \ref{algo: FeaturesExtraction} in Section \ref{subsec:PCA}}
    
    $score^{i} \leftarrow F_{\widehat{\Theta}}(\varepsilon^{i})$\;
    
    $\hat{A}^i \leftarrow \mathds{1}_{score^{i}>\hat{s}} $\;
    
    \SetKwInOut{Output}{outputs}
    \Output{identification label $\hat{A}^i$.} \end{algorithm}

     \tikzset{%
  every neuron/.style={
    circle,
    draw,
    minimum size=1cm
  },
  neuron missing/.style={
    draw=none, 
    scale=2,
    text height=0.333cm,
    execute at begin node=\color{black}$\vdots$
  },
}













\subsection{Anomaly Localization Step}
\label{subsec:Anomalylocalization}

Once the time series containing an anomaly have been identified, the second step of our approach aims at localizing an abnormal observation among each contaminated time series. Again we use the reconstruction errors defined in \eqref{eq:recerror} as inputs of this second step. The difference lies on the transformation these reconstruction errors undergo before labelling the different time stamp observations.
Namely, we now consider the following transformation of the reconstruction errors:
\begin{equation*}
    F(\varepsilon^{i}) = \lvert \varepsilon^i \rvert = \lvert X^{i} - \widehat{X}^{i}  \rvert,
\end{equation*} 
with $\lvert \cdot \rvert$ meant component wise. The model input $F(\varepsilon^{i}) \in \mathbb{R}^{p}_{+}$ is thus assigned to each time series $X^{i} \in \mathbb{R}^{p}$. The time stamp of occurrence of the anomaly is given by 
\begin{align*}
    \widehat{L}^{i} := \arg\max_{j} \left\{F(\varepsilon^{i}_{t_j}) : j \in \{ 1,\ldots, p \}\right\}.
\end{align*}
Algorithm \ref{algo: localizationModel} recaps the anomaly localization model.  

    \begin{algorithm}[H]
     \SetAlgoLined
     \caption{Anomaly Localization Model}
    \label{algo: localizationModel}
     \SetKwInOut{Require}{inputs}
     \Require{time series $X^i$ to analyze.}
    $\varepsilon^{i} \leftarrow \mathrm{PCAFeaturesExtraction}(X^i)$
    
    $\widehat{L}^i \leftarrow \arg\max_{\jmath} \left\{\lvert \varepsilon^{i}_{t_\jmath}\rvert : \jmath \in \{ 1,\ldots, p \}\right\}\;$
    
     \SetKwInOut{Output}{outputs}
     \Output{anomaly location $\widehat{L}^i$.} 
     \end{algorithm}

Note that anomalies are not necessarily extrema. For this reason, the above PCA features extraction step is necessary. Building on the reconstruction errors, the model identifies these extrema-anomalies, but also abnormal observations which are not necessarily extrema. This subtlety of the nature of the observations underlines the importance of going further than taking the index of the highest observed value of the contaminated time series $X^i$.

\tikzstyle{decision} = [diamond, draw, fill=gray!20, 
    text width=4.5em, text badly centered, node distance=4cm, inner sep=0pt]
\tikzstyle{output} = [coordinate]

\tikzstyle{corner} = [rectangle, draw, fill=white!20, 
    text width=2em, text centered, rounded corners, minimum height=0.1em, minimum width=0.1em,node distance=3cm]
\tikzstyle{block} = [rectangle, draw, fill=gray!20, 
    text width=6em, text centered, rounded corners, minimum height=4em,node distance=3cm]
\tikzstyle{blocktransp} = [rectangle, draw,
    text width=6em,  rounded corners,minimum width=5cm, minimum height=1cm,node distance=3cm]
\tikzstyle{blockEnd} = [rectangle, draw, fill=gray!20, 
    text width=2em, text centered, rounded corners, minimum height=0.5em, minimum width=0.25em,node distance=3cm]
\tikzstyle{line} = [draw, -latex']
\tikzstyle{linecust} = [draw]
\tikzstyle{cloud} = [draw, ellipse,fill=red!40, node distance=3cm,
    minimum height=2em]
    
\def\Popsize{{\text{Pop}_{size}}}\def\Popsize{P}

\begin{figure}
    \centering
    \begin{tikzpicture}[node distance = 2cm,auto]
    \node [block] (step1) {Features extraction};
    \node [cloud, left of=step1] (input) {$X$};
    \node [cloud, right of= step1] (output) {$\varepsilon$};
    \node [block, below of= step1] (stepad1) {Identification};
    \node [cloud, left of= stepad1] (inputad1) {$\varepsilon$};
    \node [cloud, right of= stepad1] (outputad1) {$\widehat{A}$};
    \node [decision, below of= stepad1] (suspornot) {Is $X$ contaminated?};
    \node [cloud, left of= suspornot](inputdec) {$\widehat{A}$};
    \node [blockEnd, right of= suspornot] (outputdecno) {End};
    \node [block, below of = suspornot] (stepad2) {Localization};
    \node [cloud, left of = stepad2] (inputad2) {$\varepsilon$};
    \node [cloud, right of = stepad2] (outputad2) {$\hat{L}$};
    \node [block, below of = stepad2] (stepad3) {Imputation};
    \node [cloud, left of = stepad3] (inputad3) {$X,\hat{L}$};
    \node [cloud, right of = stepad3] (outputad3) {$\widetilde{X}$};
    \node [output, below of= outputad3] (endModel) {};
    \node [output, below of= inputad3] (endModel2) {};
    \node [output, left of= endModel2] (endModel3) {};
    \node [output, left of= input] (endModel4) {};

    \path [line,dashed] (input) -- (step1);
    \path [line,dashed] (step1) -- (output) ;
    \path [line](step1)--(stepad1);
    \path [line,dashed] (inputad1) -- (stepad1);
    \path [line,dashed] (stepad1) -- (outputad1);
    \path [line] (stepad1) -- (suspornot);
    \path [line,dashed] (inputdec) --(suspornot); 
    \path [line, dashed] (suspornot) -- node {no} (outputdecno);
    \path [line,dashed] (suspornot) -- node {yes} (stepad2);
    \path [line,dashed] (inputad2) -- (stepad2);
    \path [line, dashed] (stepad2) -- (outputad2);
    \path [line] (stepad2) -- (stepad3);
    \path [line,dashed] (inputad3) -- (stepad3);
    \path [line, dashed] (stepad3) -- (outputad3);
    \draw (outputad3) -- (endModel);
    
    \draw (endModel) -- (endModel2);
    \draw (endModel2) -- (endModel3);
    \draw (endModel3) -- (endModel4);
    \path [line] (endModel4) -- (input);
    
\end{tikzpicture}
    \caption{Flow chart of our two step anomaly detection model (PCA NN), depicting the process a time series $X$ goes through. }
    \label{fig:FlowChartModel}
\end{figure}
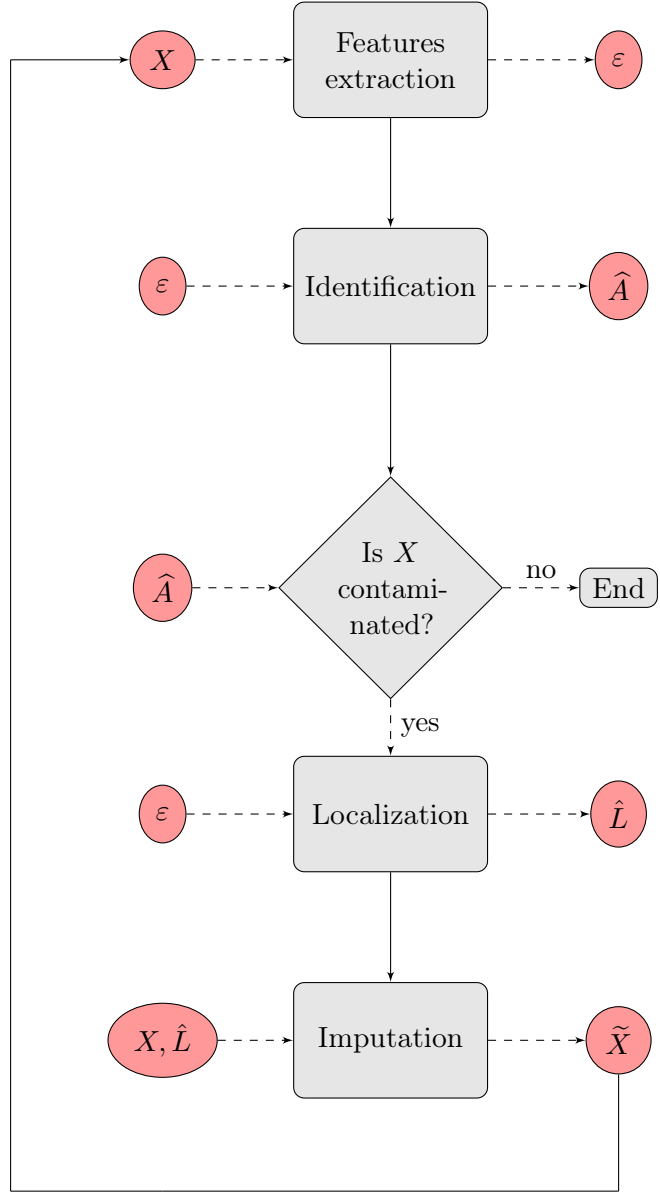

The flow chart of Figure \ref{fig:FlowChartModel} along with Algorithm \ref{algo: AnomalyDetectionModel} summarize our approach. When a time series $X$ is given to our model, we suggest a new representation of $X$, namely $\varepsilon$, through a features engineering step involving PCA. The resulting representation $\varepsilon^i$ feeds the first component of the model, i.e. the identification step, which evaluates the time series propensity of being contaminated. The optimal parameter $\widehat{\Theta}$ of the identification model solves

\[
\widehat{\Theta} = \arg \min_{\Theta=\left(\bm{W},\bm{b},s\right)} \sum_{X\in\bm{X}} \mathcal{L}_{\Theta} \left( A, \hat{A}\right), 
\]
where $\hat{A} = \mathds{1}_{F\left((\Omega\Omega^\top - I_{p})X\right)>s}$ and $\mathcal{L}_{\Theta}$ is the loss function defined in \eqref{eq:myLoss}. Then, if the model considers the time series as contaminated, the localization model takes over to localize the abnormal value in $X$. For this second step of the model, the time stamp of occurrence of the anomaly is given by 
\[
\arg\max_{j}  \quad  \left\{\left( X \left(\Omega^\top\Omega - I_{p}\right) \right)_j : j \in \{ 1,\ldots, p \}\right\}
\]

Finally, the anomaly is imputed. As our approach integrates the computation of a reconstruction of the time series, we could replace the anomaly with the corresponding reconstructed value. The model will be then able to detect the anomaly and suggest an imputation value. However, numerical tests described in Section \ref{subsection:imputation} show that imputations with naive approaches perform better.

    \begin{algorithm}[H]
 
     \SetAlgoLined
     \caption{Anomaly Detection Model }
    \label{algo: AnomalyDetectionModel}
     \SetKwInOut{Require}{inputs}
     \Require{time series $X^i$ to analyze, \\calibrated identification model parameter $\widehat{\Theta}$} 

     $\widehat{A}^i \leftarrow \mathrm{IdentificationModel}(X^i,\widehat{\Theta})$  \tcp*[f]{Algorithm \ref{algo: IdentificationModel}} \\
     \If{$\widehat{A}^i=1$}{
     $\widehat{L}^i \leftarrow \mathrm{LocalizationModel}(X^i)$ \tcp*[f]{Algorithm \ref{algo: localizationModel}}
     }
         \SetKwInOut{Output}{outputs}
     \Output{identification label $\widehat{A}^i$,}
     \hspace{42pt} anomaly localization $\widehat{L}^i$.
     \end{algorithm}

\section{Data Generation Process}\label{section: data}

Since anomalies are rare by definition, real data sets are very imbalanced: the proportion of anomalies compare to normal observations is very small, making the learning phase of the model difficult. This has led us to first consider synthetic data for the model calibration. The corresponding data set is obtained through a three-step process including time series simulations, contamination and data augmentation.  We point out that, in the data simulation, care was taken to ensure that the generated data sets stay realistic. In particular, only few anomalies were added to time series as described in Section \ref{subsec:DataContamination}. To this extent, we still face the problem of scarcity of anomalies in this synthetic framework, and we provide some preprocessing steps to sidestep this issue as well.

\subsection{Data Simulation}
\label{subsection:datasim}
In this section, we describe the model used to simulate the data. Recall that our primary motivation is to detect anomalies in financial time series. For that purpose, we consider share price sample paths generated through the Black and Scholes model, i.e. geometric Brownian motions. Under this framework, the share price $S_t$ is defined by
\begin{align}
    S_t = S_0  \exp\left(\left(\mu - \frac{1}{2}\sigma^2\right)t + \sigma W_t\right).
    \label{eq:diffeq}
\end{align}
where $W$ represents a standard Brownian motion, $\mu$ the drift, $\sigma$ the volatility of the stock and $S_0$ the initial stock price.

Let $N$ stocks $S^{1}, \ldots, S^{N}$ simulated simultaneously from this model, where $S^{i}:= \left\{S_{t_0}^{i},S_{t_1}^{i},\dots,S_{t_T}^{i}\right\}$ is the time series of length $T$ representing the (time discretized) path diffusion of the $i$-th stock. Each stock $S^i$ has its own drift $\mu^i$, volatility $\sigma^i$ and initial value $S^i_{t_0}$. The paths parameters are selected randomly according to 

\begin{align}
   S_{t_0}^i \sim \mathcal{N}\left(100,1\right), \quad \quad    \mu^i \sim \mathcal{U}\left([0.01,0.2]\right), \quad \quad   \sigma^i \sim \mathcal{U}\left([0.01,0.1]\right), \quad \, i=1,\ldots, N . 
   \label{eq:genparams}
\end{align}

For the sake of realism, the Brownian motions driving the  $N$-stocks are correlated.
The contamination of the obtained time series by anomalies is described in the next section.

\subsection{Time Series Contamination}
\label{subsec:DataContamination}

We apply a quite naive approach to introduce anomalies into our time series. We introduce the same fixed number of anomalies $n^{anom}$ to each time series by applying a shock on some original values of the observed time series. Formally, the  $\jmath$-th added anomaly is characterized by its location $t_{\jmath}$ corresponding to the time stamp at which the anomaly has occurred, its shock $\delta_{\jmath}$, the amplitude of the shock is given by $|\delta_{\jmath}|$ and its sign by $\mathrm{sgn}\left(\delta_{\jmath}\right)$. We denote by $S^{a,i}$ the time series resulting from the contamination of the $i$-th clean time series $S^i$. For $i \in \{1,\dots ,N \}$, let $\mathcal{J}^i$ be the set of indices of the time stamps at which an anomaly occurs for the $i$-th time series. For $\jmath \in \mathcal{J}^i$, the abnormal values are
\begin{align}
     S_{t_{\jmath}}^{a,i} = S_{t_{\jmath}}^{i} \left( 1 + \delta_{\jmath}\right).
     \label{eq:contam}
\end{align}

The location, sign and amplitude of the shocks are generated randomly according to uniform distributions:
\begin{align*}
   \mathcal{J} \sim \mathcal{U}_{N,n^{anom}}\left(\{1,\ldots,p\}\right), \quad \quad    \mathrm{sign}\left(\delta\right) \sim \mathcal{U}_{N,n^{anom}}\left(\{-1,1\}\right), \quad \quad   \lvert \delta\rvert \sim \mathcal{U}_{N,n^{anom}}\left([0,\rho]\right),
   \label{eq:contamParams}
\end{align*}
where $\rho$ is an upper bound on the shock amplitude.

We define the anomaly mask matrix $\mathbb{R}^{N \times T}$ by setting, for $i=1,\ldots,N$, and $j=1,\ldots, T$, 

\begin{equation}
\label{eq:anomalymask}
  \mathcal{T}_{i,j} = \begin{cases}
     1+ \delta_j   & \text{if $j \in \mathcal{J}^i$.} \\
     1 & \text{otherwise.}
\end{cases}
\end{equation}

Under this framework, when we incorporate the anomalies to the clean time series driven by the geometric Brownian motion, we assign labels to the time series observations according to whether the values correspond to anomalies or normal observations.  We thus provide the labels $Y^{i}_{t_j} $ associated with each value $S_{t_j}^{a,i}$. Hence, for $i=1,\ldots,N$, and $j=1,\ldots, T$,

\begin{align}
    Y^i_{t_j} = \begin{cases}
       1  & \text{if $j \in \mathcal{J}^i$.} \\
     0 & \text{otherwise.}
\end{cases}
\label{eq:getlabels}
\end{align}

Algorithm \ref{algo: DataContamination} recaps the time series contamination procedure, where AnomalyMask and GetLabels refer to the operators defined by \eqref{eq:anomalymask} and \eqref{eq:getlabels}.

\begin{algorithm}[H]
     
     \SetAlgoLined
     \caption{Data Contamination }
     \label{algo: DataContamination}
     \SetKwInOut{Require}{inputs}
     \Require{amplitude range $\rho$,\\ number of anomalies to add in time series $n^{anom}$,\\ data $\bm{S}$}
     
     $\mathrm{sgn}\left(\delta\right)
     \leftarrow \mathcal{U}_{N,n^{anom}}\left(\{-1,1\}\right)$\\
     $ \lvert\delta\rvert
     \leftarrow \mathcal{U}_{N,n^{anom}}\left(\{0,\rho\}\right)$\\
     $ \mathcal{J} \leftarrow \mathcal{U}_{N,n^{anom}}\left(\{1,\ldots,p\}\right)$\\
     $\mathbb{R}^{N \times T} \ni\mathcal{T} \leftarrow  \mathrm{AnomalyMask}(\mathcal{J},\delta)$\\
     $\bm{S}^{a} \leftarrow \bm{S} \circ \mathcal{T}$\\
     $Y \leftarrow \mathrm{GetLabels}\left(\mathcal{J}\right)$

    \SetKwInOut{Output}{outputs}
     \Output{time series with anomalies $\bm{S}^{a}$,}
     \hspace{42pt} labels associated with each value of the time series $\bm{Y}$.
     \end{algorithm}

\subsection{Data Augmentation}


By definition, anomalies are rare events and thus represent only a low fraction of the data set. Yet the suggested approach needs an important training set for an efficient learning. To overcome this issue, we apply a sliding window data augmentation technique \citep{le2016data}. This method not only extends the number of anomalies within the data set, but also allows the model to learn that anomalies could be located anywhere in the time series. It consists in extracting $N_{p} = T- p + 1$ sub-time series of length $p$ of the initial observed time series of length $T$.

Note that we have to split the data into train and test sets before augmentation to guarantee that the time series considered in the training set do not share any observation with the ones we use for the model evaluation. In view of simplification, we introduce the data augmentation process, without loss of generality, for $\bm{S}$ and $\bm{Y}$. In practice this process has to be applied to $\bm{S^{Train}}$,$\bm{Y^{Train}}$ and $\bm{S^{Test}}$, $\bm{Y^{Test}}$ separately.  

For $i=1,\ldots, N$ and $q \in \{1,\ldots,N_{p}\}$, the sub-time series $S^{i,q}$ and the associated labels $Y^{i,q}$ are defined by (see Algorithm \ref{algo: Slidding})
\begin{align*}
   S^{i,q} &= \left\{S_{t_j}^{i} \, , \ j =q,\dots,q + p - 1  \right\} \\
   Y^{i,q} &= \left\{Y_{t_j}^{i} \, , \  j =q,\dots,q + p - 1  \right\} .
\end{align*}
Each sub-time series $S^{i,q+1}$ thus results from the shift forward in time of one observation of the previous sub-time series $S^{i,q}$.  The final data set $\bm{X}$ and the labels $\bm{{}^{s}Y}$ are then defined as the matrices whose rows correspond to the sub-time series $S^{i,q}_{q=1,\ldots, N_p; i= 1,\ldots,N}$, and $Y^{i,q}_{q=1,\ldots, N_p; i= 1,\ldots,N}$, respectively,  i.e.
\begin{align*}
    \bm{X} &= \left(S^{1,1}, S^{1,2}, \ldots , S^{1,N_{p}}, S^{2,1} , \ldots , S^{i,q}, \ldots, S^{N,N_{p}}
\right)^{\top}, \\
\bm{{}^sY} &= 
\left(
Y^{1,1}, Y^{1,2}, \ldots , Y^{1,N_{p}}, Y^{2,1} , \ldots , Y^{i,q}, \ldots, Y^{N,N_{p}}
\right)^{\top}.
\end{align*}

With this data configuration, an observation refers to a time series $S^{i,q}$ obtained through the sliding window technique. Two observations may represent the same stock but on different time intervals.
\begin{algorithm}[H]
     
     \SetAlgoLined
     \caption{Data Augmentation with sliding window technique}
    \label{algo: Slidding}
     \SetKwInOut{Require}{inputs}
     \Require{window size $p$,}
     \hspace{42pt} data and labels $\bm{S}$,$\bm{Y}$; \\
     Initialize empty slided time series and labels matrices $\bm{X}$, $\bm{{}^sY}$;\\
     \For{i :=1 to N}{
         \For{q :=1 to $N_{p}$}{
         $S^{i,q} \leftarrow \left\{S_{t_j}^i | j \in \{q,\dots,q+p-1\}\right\}$\\
         
         $Y^{i,q} \leftarrow \left\{Y_{t_j}^i | j \in \{q,\dots,q+p-1\}\right\}$\\
         
         }
         $\bm{X} \leftarrow \mathrm{Concatenate}\left(\bm{X}, S^{i,q}\right)$
         
         $\bm{{}^sY}\leftarrow \mathrm{Concatenate}\left(\bm{{}^sY}, Y^{i,q}\right)$\\
         
     }
    \SetKwInOut{Output}{outputs}
     \Output{resulting slided time series and labels $\bm{X}, \bm{{}^sY}$} 
     \end{algorithm} 
While the fact that the observations share the same values may be argued to wrongly impact the learning process, we point out that a real benefit can be drawn from this situation. Indeed, thanks to this sliding window technique, the number of anomalies is considerably increased. This technique also allows the model to learn that anomalies could be located anywhere in the time series, reducing the dependency on the event location \citep{um2017data}. 

However, once we apply the sliding window technique, we do not only extend the number of contaminated time series: the number of uncontaminated time series is also increased. But the minority class (contaminated time series) has, at least, a significant number of instances, denoted by $N^c$. In order to get a more balanced data set for the training set, we perform an undersampling, selecting randomly $N^c$ observations from the $N^u$ uncontaminated time series without any anomalies (RandomSampling in Algorithm \ref{algo: Pruning}). The resulting retained number of observations $2N^c$ is now sufficient to train the model. The test set in turn is imbalanced. To sharpen the imbalanced characteristic of the test set, we specify a contamination rate $r_c$, which corresponds to the rate of contaminated time series in the data set. The construction of the test set is described in Algorithm \ref{algo: Pruning}. 

    \begin{algorithm}[H]
     
     \SetAlgoLined
     \caption{Time series selection }
    \label{algo: Pruning}
     \SetKwInOut{Require}{Inputs}
     \Require{slided data and labels $\bm{X}$,$\bm{{}^sY}$,}
     \hspace{42pt} contamination rate $r_c$ (for test set) \\
     \medskip
 
         $\bm{X},\bm{{}^sY} \leftarrow \bm{X}\mathds{1}_{\mathrm{sum}(\bm{{}^sY})\leq 1},\bm{{}^sY}\mathds{1}_{\mathrm{sum}(\bm{{}^sY})\leq 1}$ \tcp*{Keep time series with at most one anomaly.}
         \If{Train set}{

         $\\ N^c = \mathrm{Card}\left(\bm{{}^sY}\mathds{1}_{\mathrm{sum}(\bm{{}^sY})= 1}\right)$\\
         \tcp*[f]{Randomly select the indexes of $N^c$ uncontaminated time series} \; \\
         $\mathrm{index}^u = \mathrm{RandomSampling}\left(\left\{i; \;  \mathrm{sum}(Y_s^i)=0, i \in \{1,\ldots,NN_p\}\right\}, N^c\right)$\\}
         
         \If{Test set}{

         $\\ N^c = \mathrm{Card}\left(\bm{{}^sY}\mathds{1}_{\mathrm{sum}(\bm{{}^sY})= 1}\right)$\\ \;
         $N^u = \left \lceil \frac{N^c(1-r_c)}{r_c}\right\rceil$\\
         \tcp*[f]{Randomly select the indexes of $N^c$ uncontaminated time series} \; \\
         $\mathrm{index}^u = \mathrm{RandomSampling}\left(\left\{i; \;  \mathrm{sum}(Y_s^i)=0, i \in \{1,\ldots,NN_p\}\right\}, N^u\right)$\\
          }

    $\\ \mathrm{index}^c =\left\{i; \;  \mathrm{sum}(Y_s^i)=1, i \in \{1,\ldots,NN_p\}\right\}$\\
 $\bm{X} \leftarrow \left(X^i, \mathrm{for} \; i \; \in \; \mathrm{index}^u \cup \mathrm{index}^c\right)$\\
 $\bm{{}^sY} \leftarrow \left(Y_s^i, \mathrm{for} \; i \; \in \; \mathrm{index}^u \cup \mathrm{index}^c\right)$\\

    \SetKwInOut{Output}{outputs}
     \Output{time series $\bm{X}$ with at most one anomaly, \\
     corresponding labels for identification task $\bm{A}$,\\
     corresponding labels for localization task $\bm{L}$.}
     \end{algorithm} 
     
As mentioned in Section \ref{subsection:modeldescription}, Algorithm \ref{algo: AnomalyDetectionModel} is designed to predict the localization of only one anomaly, (if there is more than one anomaly in the time series, it should be run iteratively as detailed in Section \ref{subsection:modeldescription}). Here we only consider time series with at most one anomaly (without loss of generality).


\begin{assumption}
A time series $S^{i,q}$ contains at most one anomaly among all its observed values.
\end{assumption}

We assign the identification label $A^{i,q}$ (see Section \ref{subsec:MethodNotation}) to each $S^{i,q}$ following the rule 

\begin{equation*}
A^{i,q} = \sum_{j=q}^{q+p -1} Y^{i}_{t_j} = \begin{cases}
       1  & \text{if there is an anomaly among the observed values of $S^{i,q}$} \\
     0 & \text{otherwise.}
     \end{cases}
\end{equation*}
Regarding the localization labels, we recall that they only concern the time series with an anomaly, therefore $L^{i,q}$ is defined following

\begin{equation}
L^{i,q} = \arg\max_{j} Y^{i,q} =\arg\max_{j} \left\{Y_{t_j}^{i} \, , \  j =q,\dots,q + p - 1  \right\}
\label{LabelFromY}
\end{equation}

With Algorithm \ref{algo: Labeling}, we give a rundown of the construction process of the identification and localization labels, namely $\bm{A}$ and $\bm{L}$ departing from $\bm{{}^sY}$.

The supervised learning framework is adopted herein, since we have at our disposal labelled data.

The resulting matrices with the observed values of the time series in $\bm{X}$ the associated  identification labels in $\bm{A}$ and localization labels in $\bm{L}$  constitute the data set used for our model calibration and evaluation.

    \begin{algorithm}[H]
     
     \SetAlgoLined
     \caption{Time series labelling}
    \label{algo: Labeling}
     \SetKwInOut{Require}{Inputs}
     \Require{slided labels $\bm{{}^sY}$;}
     
         $\bm{A} \leftarrow \mathrm{sum}(\bm{{}^sY})$ \; \\  
        
        $\bm{L} \leftarrow \arg\max \left\{\bm{{}^sY}\mathds{1}_{\mathrm{sum}(\bm{{}^sY}) = 1}\right\}$\\

    \SetKwInOut{Output}{outputs}
     \Output{corresponding labels for identification task $\bm{A}$,\\
     corresponding labels for localization task $\bm{L}$.}
     \end{algorithm}


\section{Model Evaluation: Setting the Stage}\label{section: model_eval}

After introducing the relevant performance metrics used to assess the PCA NN performance, we describe the synthetic data used for the numerical experiments of Sections \ref{section: model_eval2} to \ref{section:VaR}. We then briefly describe the process of the latent space dimension calibration.


\subsection{Performance Metrics}
\label{subsection:perfmetrics}
Common methods to assess the performance of binary classifiers include true positive and true negative rates, as well as ROC (Receiver Operating Characteristics) curves, displaying the true positive rate against the false positive rate. These methods, however, are uninformative when the classes are severely imbalanced.  In this context, $F_1$-score and  Precision-Recall curves (PRC) should be used \citep{brownlee2020imbalanced,prc}. They are both based on the values of
\[
\mathrm{Precision} (s)= \frac{\text{true positives}}{\text{true positives} + \text{false positives}}
\]
against the values of
\[
\mathrm{Recall} (s) = \frac{\text{true positives}}{\text{true positives} + \text{false negatives}} \, 
\]
where $s$ is a cut-off probability varying between $0$ and $1$. 
Precision quantifies the number of correct positive predictions out all positive predictions made, while Recall (often also called Sensitivity) quantifies the number of correct positive predictions out of all positive predictions that could have been made. Both focus on the Positives class (the minority class, anomalies) and disregard the Negatives (the majority class, normal observations). 

Our $F_1$-score \citep{fmes1, van1979information} combines these two measures in a single index defined as 
\begin{equation}
  F_1 = 2\cdot \frac{ \mathrm{Precision}\cdot \mathrm{Recall}}{\mathrm{Precision}+\mathrm{Recall}}.
\end{equation}
The closer the $F_1$-score to 1, the better the prediction model.  

A PRC displays the values of Precision and Recall as the cut-off $s$ varies from $0$ to $1$. The PRC of a skillful model bows towards the point with coordinates $(1,1)$. The curve of a no-skill classifier is an horizontal line at some $y$-level proportional to the proportion of Positives in the data set. For a balanced data set this proportion is just 0.5 \citep{brownlee2020imbalanced}.

PRC and $F_1$-score are complementary in our approach. The $F_1$-score is used on the anomaly scores outcomes of the models, to identify the best configuration and the best model. A PRC is used to select the best cut-off value used in the prediction of the two classes, for each model.










\subsection{Synthetic Data Set}

The anomaly detection task is performed on $N=20$ stocks simultaneously. The stock prices are diffused according to the Black and Scholes model. Each stock has its own drift and volatility and the $20$ stocks are correlated, as described in Section \ref{section: data}. Each time series represents $T=1,500$ daily stock prices, split into two sets. The first $1,000$ observations, corresponding to the train set, are used to learn the model parameters, i.e. the PCA transfer matrix and the NN weights and biases. The last $500$ observations, corresponding to the test set, are used to assess the quality of the estimated parameters when applied to unseen samples. For the application of the sliding window technique, we set the length of the resulting time series $p$ to be 206. \Table{\ref{tab:compo}} sums up the composition of each data set before and after data augmentation. 
 
\begin{table}[h!]

  \begin{center}
\begin{tabular}{c}
    \begin{tabular}{l|c|c|c} 
      & Nb of Time Series & Nb of  Observed values per time series & Nb  of  anomalies \\
      \hline
      Train set & 20 & 1,000& $4 \times 20$\\
      Test set & 20& 500 & $2 \times 20$
    \end{tabular}\\~\\
 
    \begin{tabular}{l|c|c|c} 
       & Nb of Time Series & Nb of  Observed values per time series & Nb  of  anomalies \\
      \hline
      Train set & 12,000 & 206&  6,000\\
      Test set & 2,500& 206 &  400\\

    \end{tabular}
    \end{tabular}
    \end{center}
   \caption{Data set composition \textit{(top)} before and \textit{(bottom)} after data augmentation.}
\label{tab:compo}
\end{table}

We recall that both steps of the approach are preceeded by a features extraction step, for which the latent space dimension $k$ needs to be calibrated. As shown by the numerical results provided in Section \ref{apx:AppendixB}, the features extraction also guarantees the stationarity of the time series used for the anomaly detection task, $\bm{\varepsilon}$.

\subsection{Calibration of the Latent Space Dimension}
\label{subsec:caliK}


When performing PCA, $k$ is determined through a scree-plot, which is the representation of the proportion of variance explained by each component. The optimal $k$ corresponds to the number of principal components explaining a given level of the variance of the original data. However, this method has its limitation as stated in \citep{linting2007nonlinear}. More importantly, it is not suitable for our approach. Indeed, in our case, the number of selected principal components must achieve a trade-off between information and noise in the latent space. If we consider a too low number of principal components, we may lose information regarding the normal observations, leading to false alarms. If a too high number of principal components is retained, we may include components representing noise, which prevents the model from detecting some anomalies. 

\begin{figure}
  \centering
  \begin{tabular}{cc}
  \includegraphics[width=0.4\textwidth]{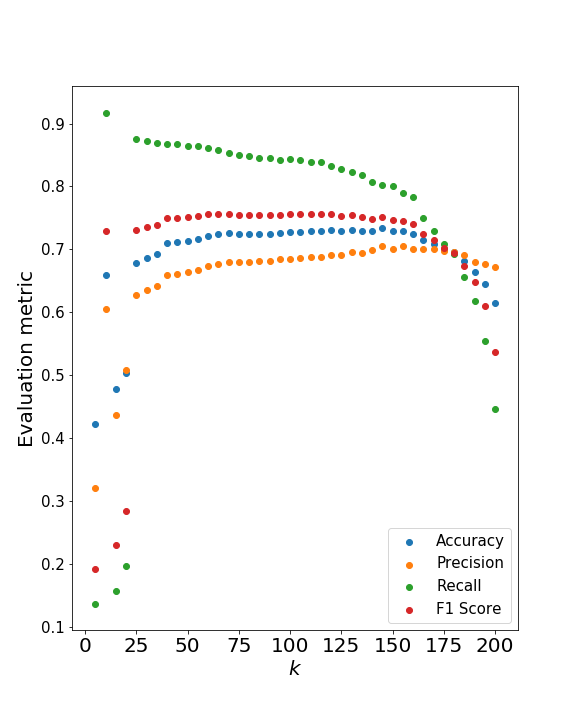} & 
  \includegraphics[width=0.4\textwidth]{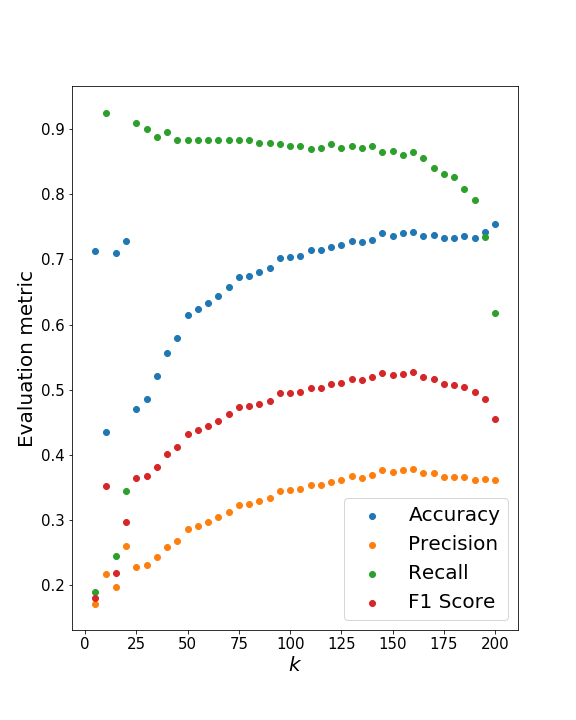}\\
  \end{tabular}
  \caption{Performance metrics on train \textit{(left)} and test \textit{(right)} sets with respect to the number of principal component $k$.}
  \label{fig:kselection}
\end{figure}

To select the optimal dimension of the latent space, we consider the distribution of anomaly scores obtained through the naive approach of Section \ref{subsec:Step1}, empirically calibrating a cut-off $\hat{s}$ through the non parametric estimation of the distribution of the anomaly scores of uncontaminated and contaminated time series. We chose to calibrate the dimension of the latent space with the naive approach, because using the NN to this end would be very costly. Hence, for each $k = 5,10,\dots,200$, we construct a PCA model from which we infer reconstruction errors which are then converted into anomaly scores. Based on these anomaly scores, we tune the cut-off value $\hat{s}$ thanks to the distributions and we finally convert the scores into labels. We evaluate the predictions of the naive approach for each value of $k$. The results on the evaluation metrics on the train set, as legitimate to choose $k$, are represented on Figure \ref{fig:kselection}. The highest values are reached for $k \in \{40,\dots,145\}$. The performance seems to be stable in terms of $F_1$-score and accuracy. Therefore, for computational reasons, we choose the optimal $k$ to be 40.

Figure \ref{fig:kselection} also shows, as expected, a downward trend of the scores when represented against the highest values of $k$. This demonstrates that when a high level of variance is explained, it become much harder to perform anomaly detection based on the reconstruction errors.

\section{Model Evaluation: Main Results}\label{section: model_eval2}

We evaluate the performance of the identification and localization stages of our approach on synthetic data, using appropriate performance metrics. We demonstrate numerically the efficiency of the PCA NN over baseline anomaly detection algorithms.  

\subsection{Contaminated Time Series Identification Step}
\label{subsection:evalstep1}

For the features extraction step, we considered a latent space dimension $k=40$. The NN built to compute the anomaly scores and convert them into labels was calibrated on the train set. The result of this calibration is shown in Table \ref{tab:ResCaliIdentification}. 
\begin{table}[H]

\centering
\begin{tabular}{l|l|l|l|l}

Data set & Accuracy &Precision & Recall  &$F_1$-score \\
\hline
Train set & 90.97 \% & 97.36\%   & 84.21\% & 90.31 \% \\
Test set  & 88.58\%  & 61.26\%   & 85.27\% & 71.30\% \\
\end{tabular}
\caption{Performance evaluation of suggested model on synthetic data set for identification step.}
\label{tab:ResCaliIdentification}
\end{table}

\begin{figure}
  \centering
  \begin{tabular}{c}
 \includegraphics[scale=0.3]{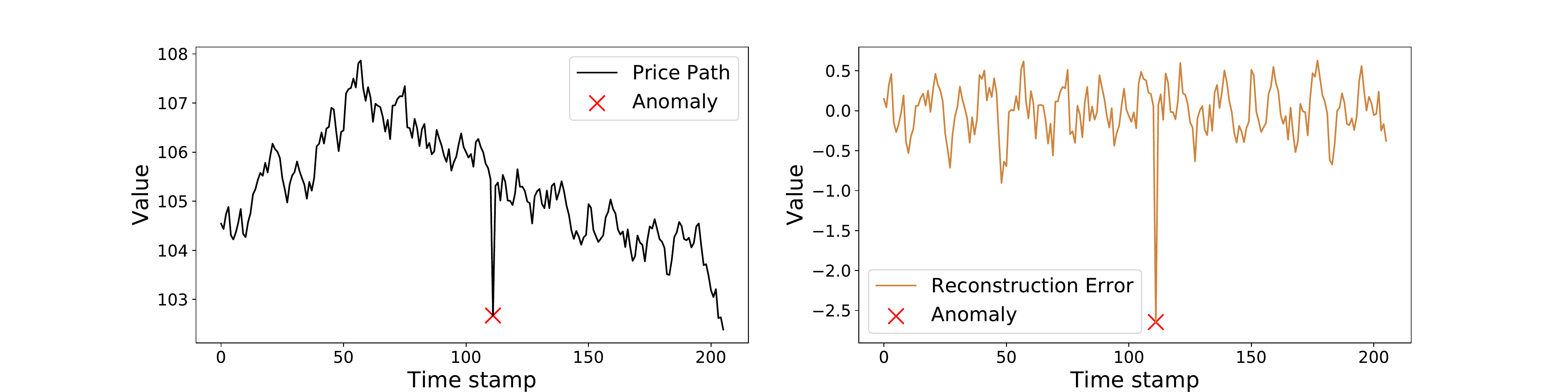}  \\
 \\
 \includegraphics[scale=0.3]{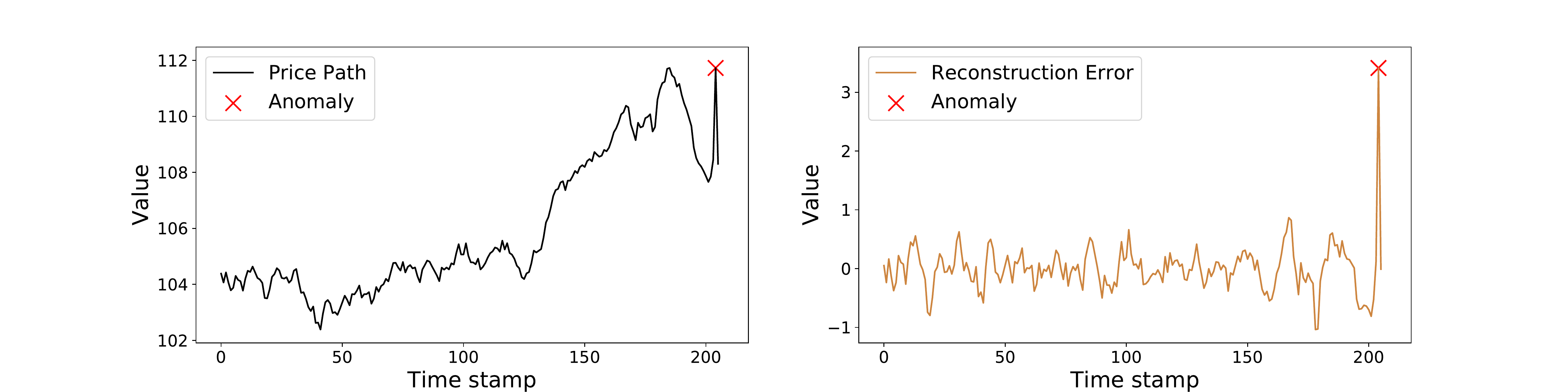}\\

  \end{tabular}
  \caption{Two examples of contaminated time series accurately identified by the model. The stock path and the reconstruction errors are represented in black and brown. The red cross shows the anomaly localization.} 
  \label{fig:detecsusp}
\end{figure}

\begin{figure}
  \centering
  \begin{tabular}{c}
  \includegraphics[scale=0.3]{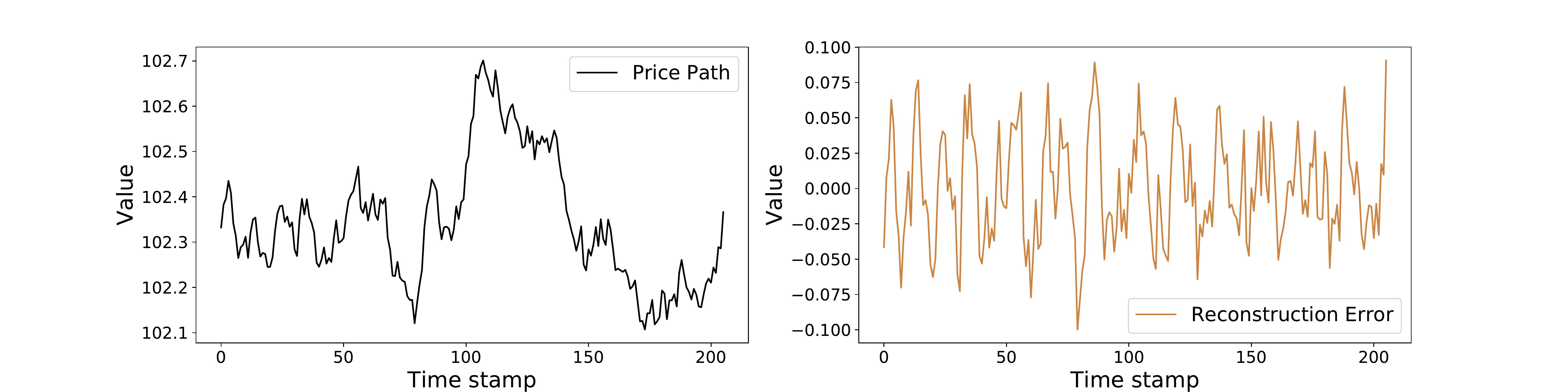}  \\
  \includegraphics[scale=0.3]{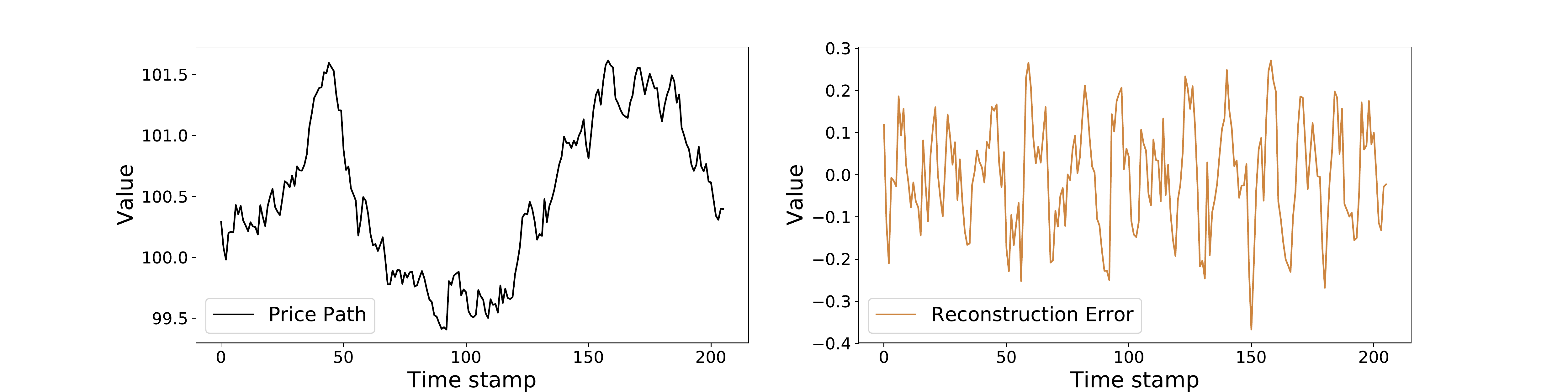} \\
  \end{tabular}
  \caption{Two examples of uncontaminated time series accurately identified by the model. The stock path and the reconstruction errors are represented in black and brown.} \label{fig:detecnon}
\end{figure}

\begin{figure}
  \centering
  \begin{tabular}{c}
 \includegraphics[scale=0.3]{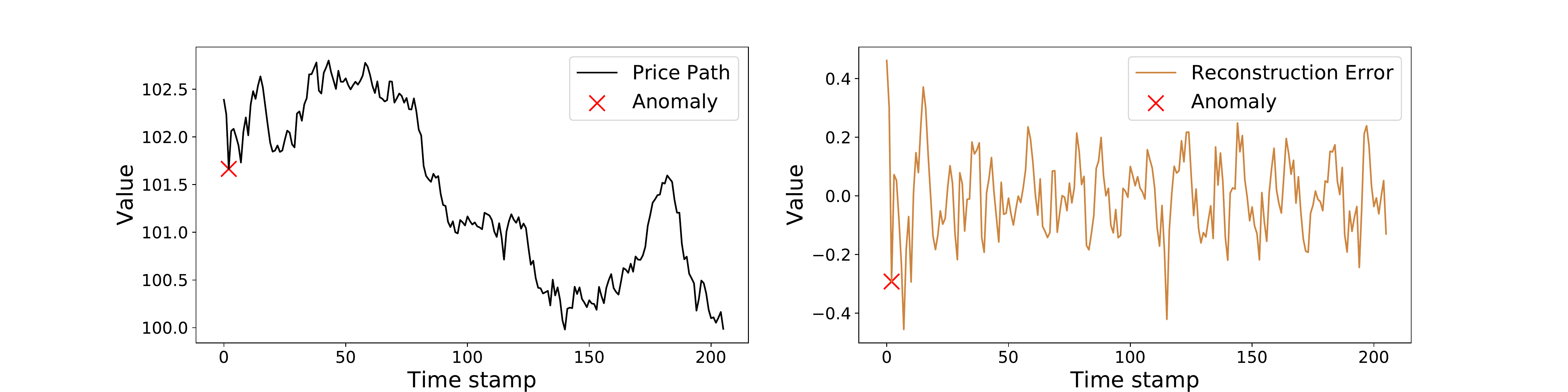}  \\
  \includegraphics[scale=0.3]{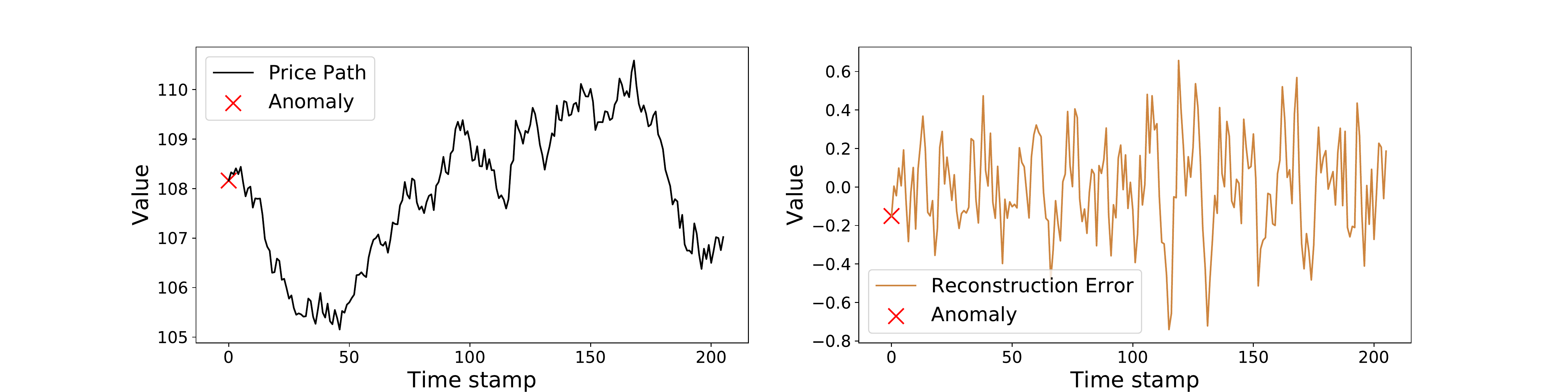} \\
  \end{tabular}
  \caption{Two examples of time series misidentified by the model. The stock path and the reconstruction errors are represented in black and brown. The red cross shows the anomaly localization.} \label{fig:detecmis}
\end{figure}

Figure \ref{fig:detecsusp} shows two contaminated time series identified as such by the model. Figure  \ref{fig:detecnon} displays two examples of time series without anomalies accurately identified by the model. Figure \ref{fig:detecmis} displays two time series misidentified by the model. 

 When an observation deviates significantly from the rest of the time series values, the model is able to recognise that the concerned time series contains an abnormal observation.

In Figure \ref{fig:detecnon} with uncontaminated time series, we see that even when there is a local upward trend in the time series values, the model is able to make the distinction between this market move and the occurrence of an anomaly. 
 
The last set of time series displays some limits of the model. The top graphs in Figure \ref{fig:detecmis} represent the situation where the model did not manage to identify the contaminated time series. One explanation could be that the size of the anomaly is not significantly large enough to be spotted by the model. Indeed, looking at the graph, the time series does not seem to contain any anomaly. In contrast, the bottom graphs in Figure \ref{fig:detecmis} show a stock price path predicted to be contaminated whereas it is not.

 We illustrate the robustness of the approach by assessing the model predictions on $100$ distinct data sets. \Table{\ref{tab:multiRunsIdent}} shows the mean and standard deviation over the multiple runs. 

\begin{table}[H]

\centering
\begin{tabular}{l|l|l|l|l}

Data set & Accuracy &Precision &Recall  &$F_1$-score \\
\hline
Train set & 79.02\% (2.4\%)& 78.69\% (2.7\%)   & 79.74\%  (4.6\%)& 79.13\% (2.7\%) \\
Test set  & 77.79\%  (3.9\%) & 41.87\% (5.2\%)   & 80.16\% (6.4\%) & 54.82\% (5.2\%) \\

\end{tabular}
\caption{Mean (standard deviation) of performance metrics of the suggested model over multiple runs for the identification step.}
\label{tab:multiRunsIdent}
\end{table}

\subsection{Anomaly Localization Step}
\label{subsection:evalstep2}
Regarding the localization step, the dummy approach  consists in taking the argmax of time series observations as the anomaly location. We distinguish between two cases, anomalies which are extrema and anomalies which are not. The quality of the detection of our model is assessed for both types of anomalies. The results are displayed in Tables {\ref{tab:ResCaliLocAll}} and \ref{tab:ResCaliLocNE}. 

\begin{table}[H]

\centering
\begin{tabular}{l|l|l|l|l}
Data set &Accuracy &Precision &Recall  &$F_1$-score \\
\hline
Train set & 89.65\% (34.87\%)& 89.81\% (40.53\%)   & 89.65\%  (34.88\%)& 89.68\% ( 36.56\%) \\
Test set & 94.39\%  (27.78\%) & 94.49\% (31.52\%)   & 94.39\% (27.78\%) & 94.38\% (28.94\%) \\
\end{tabular}
\caption{Performance evaluation of the suggested model and (the dummy approach) on synthetic data set for the localization step on all type of anomalies.}
\label{tab:ResCaliLocAll}
\end{table}

\begin{table}[H]

\centering
\begin{tabular}{l|l|l|l|l}
Data set &Accuracy &Precision & Recall  &$F_1$-score \\
\hline

Train set & 84.22\% (0\%)& 84.55\% (0\%)   & 84.28\%  (0\%)& 89.68\% (0\%) \\
Test set  & 91.92\%  (0\%) & 92.10\% (0\%)   & 91.92\% (0\%) & 91.90\% (0\%) \\
\end{tabular}
\caption{Performance evaluation of the suggested model and the dummy approach on synthetic data set for the localization step on non extrema anomalies.}
\label{tab:ResCaliLocNE}
\end{table}

\begin{figure}
  \centering
  \begin{tabular}{cc}
  \includegraphics[scale=0.3]{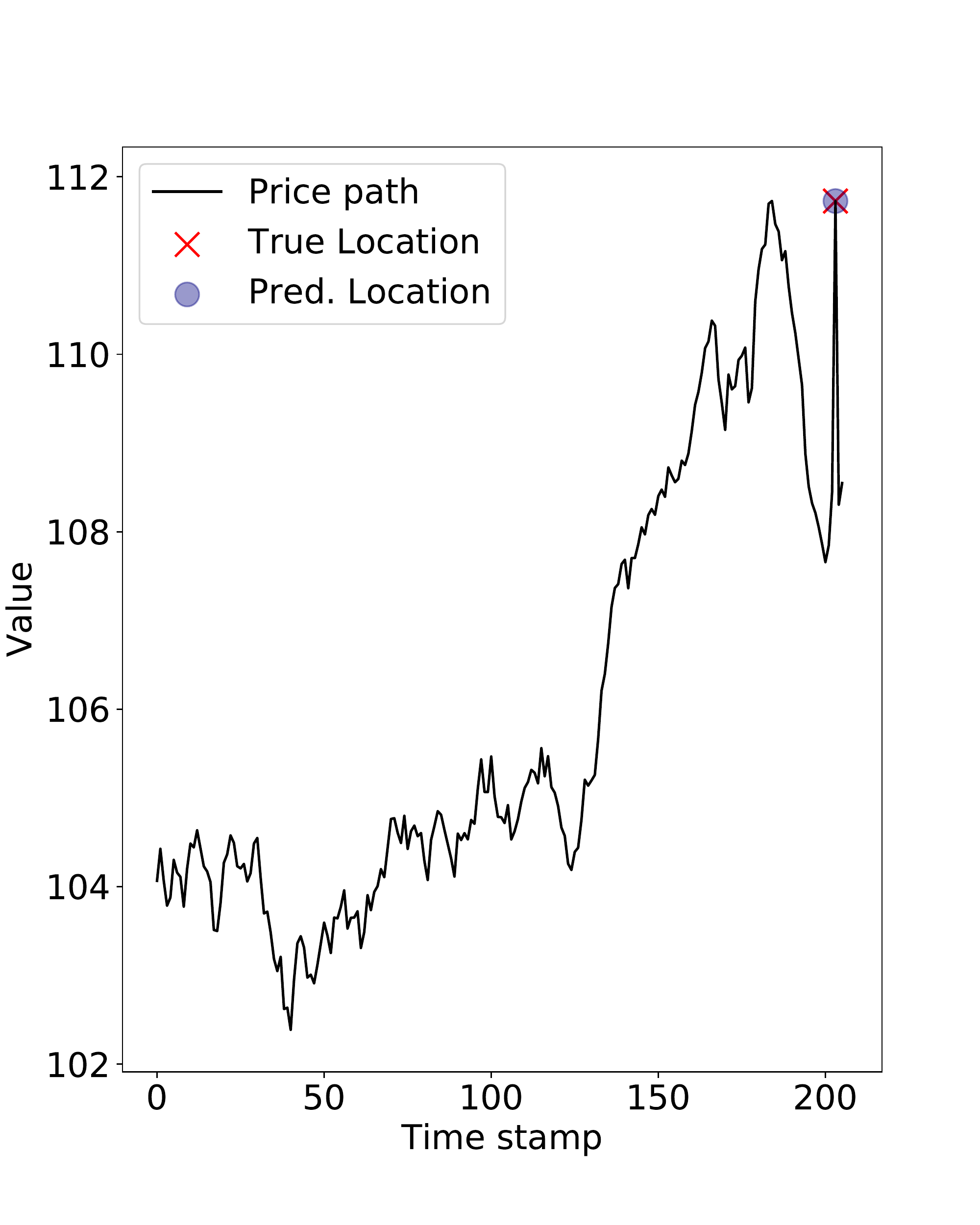} & 
 \includegraphics[scale=0.3]{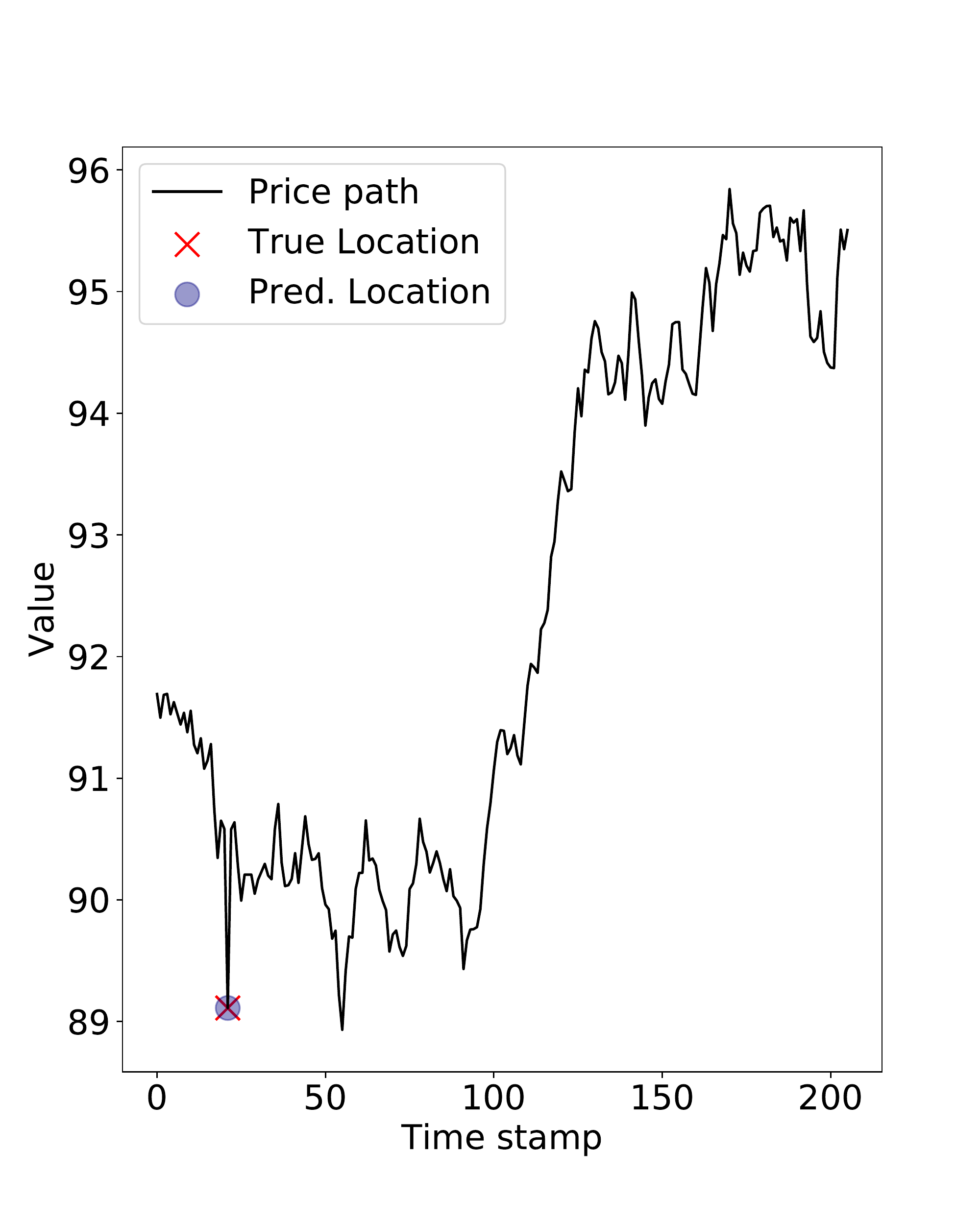}\\
 \end{tabular}
  \caption{Anomaly localization prediction on two distinct time series.} \label{fig:locavisu}
\end{figure}

 Numerical tests show the necessity of the features extraction step for the anomaly localization task, as applying the dummy approach alone is not enough when the anomaly is not an extreme value. Figure \ref{fig:locavisu} represents a stock price time series with the true and predicted anomaly location. In these examples, the locations are accurately predicted. 

To guarantee the robustness of the model on the anomaly location, we evaluate the model prediction on $100$ data sets. \Table{\ref{tab:multiRunsLocALL}} shows the mean and standard deviation for the localization step of all types of anomaly. 

\begin{table}[H]
\centering
\begin{tabular}{c|c|c|c|c}

Data set & Accuracy & Precision & Recall  &$F_1$-score \\
\hline
Train set & 89.58\% (4.3\%)& 89.58\% (4.3\%)   & 89.96\% (4.1\%)& 89.67\% (4.3\%) \\
Test set  & 89.49\%  (4.8\%) & 89.49\% (4.8\%)   & 90.00\% (4.5\%) & 89.58\% (4.7\%) 
\end{tabular}
\caption{Mean (standard deviation) of performance metrics of suggested model over multiple runs for localization step on all type of anomalies.}
\label{tab:multiRunsLocALL}
\end{table}

Table \ref{tab:multiRunsLocNE} displays the results on the localization of non extrema anomalies. These results show the importance of the features extraction step. If we only consider the maximal observed value of the time series to be the anomaly, we would not be able to localize any non-extrema anomaly. Applying Algorithm \ref{algo: localizationModel} instead leads to satisfying anomaly detection rates.

\begin{table}[H]

\centering
\begin{tabular}{l|l|l|l|l}
Data set & Accuracy & Precision & Recall  &$F_1$-score \\
\hline
Train set & 85.23\% (6.0\%)& 85.23\% (6.0\%)   & 85.87\%  (5.7\%)& 85.36\% (6.0\%) \\
Test set  & 85.17\%  (7.1\%) & 85.17\% (7.1\%)   & 86.04\% (6.8\%) & 85.30\% (7.1\%) \\
\end{tabular}
\caption{Mean (standard deviation) of performance metrics of suggested model over multiple runs for localization step on non-extreme anomalies.}
\label{tab:multiRunsLocNE}
\end{table}

\subsection{Numerical Results Against Benchmark Models}\label{subsection: benchmarking}

To assess the performance, the suggested approach is compared numerically with well-known machine learning algorithms for anomaly detection, that is isolation forest (IF), local outlier factor (LOF), density based clustering of applications with noise (DBSCAN), k-nearest neighbors (KNN), and support vector machine (SVM), all reviewed in Section \ref{apx:AppendixA}. In addition to these state of the art techniques, we consider the recent anomaly detection technique proposed by \citet{akyildirim2022applications}, referred to as sig-IF, combining a features extraction step through path signatures computation with IF.

Tables \ref{tab:perfmethodsIdentification} and \ref{tab:perfmethodsLocation} summarize the performance on the train and test sets of Section \ref{section: data} of each model, for both the contaminated time series identification and the anomaly localization steps.

\begin{table}[H]
\centering
\begin{tabular}{l|cc|cc}

& \multicolumn{2}{c|}{Train} & \multicolumn{2}{c}{Test}  \\
\hline
Model& Accuracy  &$F_1$-score& Accuracy &$F_1$-score             \\
\hline
IF          &42.31\% & 42.31\%  & 69.64\% & 7.022\% \\
LOF            &59.95\% & 59.95\% & 90.00\%  &  	62.41\% \\
DBSCAN         & 50.00\% & 66.67\% & 16.65\% & 28.53\% \\ 
sig-IF &49.48\%& 49.23\%& 72.32\% & 15.19\% \\\hline
KNN          &94.68\% & 94.39\% &64.82\% &26.69\% \\
SVM        & 81.96\% & 82.33\% & 44.39\% &27.44\% \\

PCA NN     &90.97\% &  90.31\%  & 88.58\% & 71.30\%   
\end{tabular}
\caption{Performance evaluation of unsupervised (upper part) and supervised (lower part) models for contaminated time series identification step. }
\label{tab:perfmethodsIdentification}
\end{table}

\begin{table}[H]
\centering
\begin{tabular}{l|ccccccc}
Algorithm & IF & LOF & DBSCAN &  KNN & SVM &sig-IF&PCA NN \\ 
\hline
Exec. Time & 0.6915  & 0.2015& 0.1275 & 1.725 & 4.572 &2.776& 0.003523     \\
\end{tabular}
\caption{Execution time in seconds for identification of contaminated time series step.}
\label{tab:extimeidentification}
\end{table}

The following paragraphs provide similar conclusions drawn from the results on the identification and localization steps (see Tables \ref{tab:perfmethodsIdentification} and \ref{tab:perfmethodsLocation}).

For the unsupervised learning methods, since there is not a proper learning step, even though Tables \ref{tab:perfmethodsIdentification} and \ref{tab:perfmethodsLocation} show the scores on the train and test set, these scores should be seen as the ones obtained by testing the models on independent data sets. For IF, LOF and DBSCAN models, the performance across the sets is not stable, as a significant difference could be observed between the scores on the train and test sets. The poor performance of unsupervised learning algorithms could be explained by the difficulty to estimate the contamination rate for IF and the LOF algorithms. For DBSCAN the poor performance is rather due to the high dimensionality of the data. As for the sig-IF approach, its performance in this specific case is not as overwhelming as when used to detect pump and dump operations in \citep{akyildirim2022applications}. This difference in performance is driven by the fact that, in our case, signatures are computed on the stock price path only, since no additional information is available to describe the price path we are analyzing. For pump/dump detection, instead, signatures are computed on a set of variables path including the price path and additional variables paths helpful in the identification of the pumps/dumps attempts. 

Regarding the supervised approaches, although KNN is outperforming the suggested approach on the training set, there is a non-negligible decrease of these scores on the test set. This may suggest an over-fitting on the training data, which makes KNN unable to generalize what it has learnt to unseen samples. The same trend is seen on the scores with our approach, however the loss is no as harsh as in the KNN case. 

Hence, according to the results, our PCA based approach seems to be the most suitable method for the problem at stake of anomaly detection on time series. Its satisfying performance in terms of accuracy and $F_1$-score, as well as its low computational cost for both steps (see Tables \ref{tab:extimeidentification} and  \ref{tab:extimelocalization}), make the approach stand out. 

\begin{table}[H]
\centering
\begin{tabular}{l|cc|cc}
& \multicolumn{2}{c|}{Train} & \multicolumn{2}{c}{Test}  \\
\hline
Model & Accuracy  & $F_1$-score & Accuracy & $F_1$-score\\
\hline     
IF             & 89.79\%  & 2.296\%   & 70.94\%  &1.611\%     \\
LOF            & 99.51\%  & 0\%   & 2.066\%  &0.9816\%      \\
DBSCAN         & N/A& NA       & NA          & NA          \\
sig-IF & N/A & N/A& N/A&N/A \\\hline
KNN     & 99.99\%   & 99.99\%  & 95.56\%   & 2.794\%     \\
SVM       & NA   & NA       & NA          & NA          \\
PCA NN        & 89.65\%   &  89.68\%  & 94.39\% & 94.38\%    
\end{tabular}
\caption{Performance evaluation of unsupervised (upper part) and supervised (lower part) models for anomaly localization step. Results for DBSCAN, sig-IF and SVM are not provided due to high computational cost.}
\label{tab:perfmethodsLocation}
\end{table}

\begin{table}[H]
\centering
\begin{tabular}{l|ccccccc}
Algorithm & IF & LOF & DBSCAN &  KNN & SVM& sig-IF &PCA NN \\ 
\hline
Exec. Time & 14.50  & 2.287&  N/A & 14.19 & N/A& N/A&0.002004     \\
\end{tabular}
\caption{Execution time in seconds for the anomaly localization step. Results for DBSCAN, sig-IF and SVM are not provided due to high computational cost.}
\label{tab:extimelocalization}
\end{table}

\section{Model Evaluation: Additional Results}\label{section: model_eval3}

To take our approach a leap beyond classical anomaly detection algorithms, we evaluate the anomalies imputation suggested by the PCA NN approach. We also test the robustness of the calibrated cut-off value and examine the sensitivity of the PCA NN performance to the amplitude of the anomaly.

\subsection{Anomalies Imputation}\label{subsection:imputation} 


To assess the imputation values suggested by our approach, we start from time series simulated without anomalies, then we randomly select a time stamp of the time series and add a noise to the corresponding value following the methodology described in Section \ref{section: data}.

The imputation value the PCA NN suggests, is the reconstructed observation. This imputation technique is compared to naive methods of missing values imputation, namely backward fill (BF), consisting in replacing the anomaly by the previous value, and linear interpolation (LI). The choice of these imputation methods is motivated by their low computational cost. In fact, the PCA NN imputation comes at no additional cost as stated before. Therefore, we only challenge its performance using methods with similar complexity. To assess the quality of the imputation value suggested by each approach, we consider the following metrics. The imputation errors are computed on each time series $i$ as 

\begin{align*}
    \mathrm{ImputationError}^{i} =  \sqrt{\sum_j \frac{\left(S^{i}_{t_j} -\widetilde{S}^i_{t_j}\right)^2}{n^{anom}}}\; ,
\end{align*}
where $\widetilde{S}^i$ refers to as the $i$-th path price with imputed values. The error on the covariance matrix is computed as  

\begin{align*}
    \mathrm{ErrCov} = \|\Sigma - \widetilde{\Sigma}\|_{\mathrm{Frob}},
\end{align*}
where $\|\cdot\|_{\mathrm{Frob}}$ is the Frobenius norm, $\Sigma$ the sample covariance matrix,  and $\widetilde{\Sigma}$ the covariance matrix estimated on the data after the anomalies have been replaced by their respective imputation values.

\begin{figure}
    \centering
    \includegraphics[width=\linewidth]{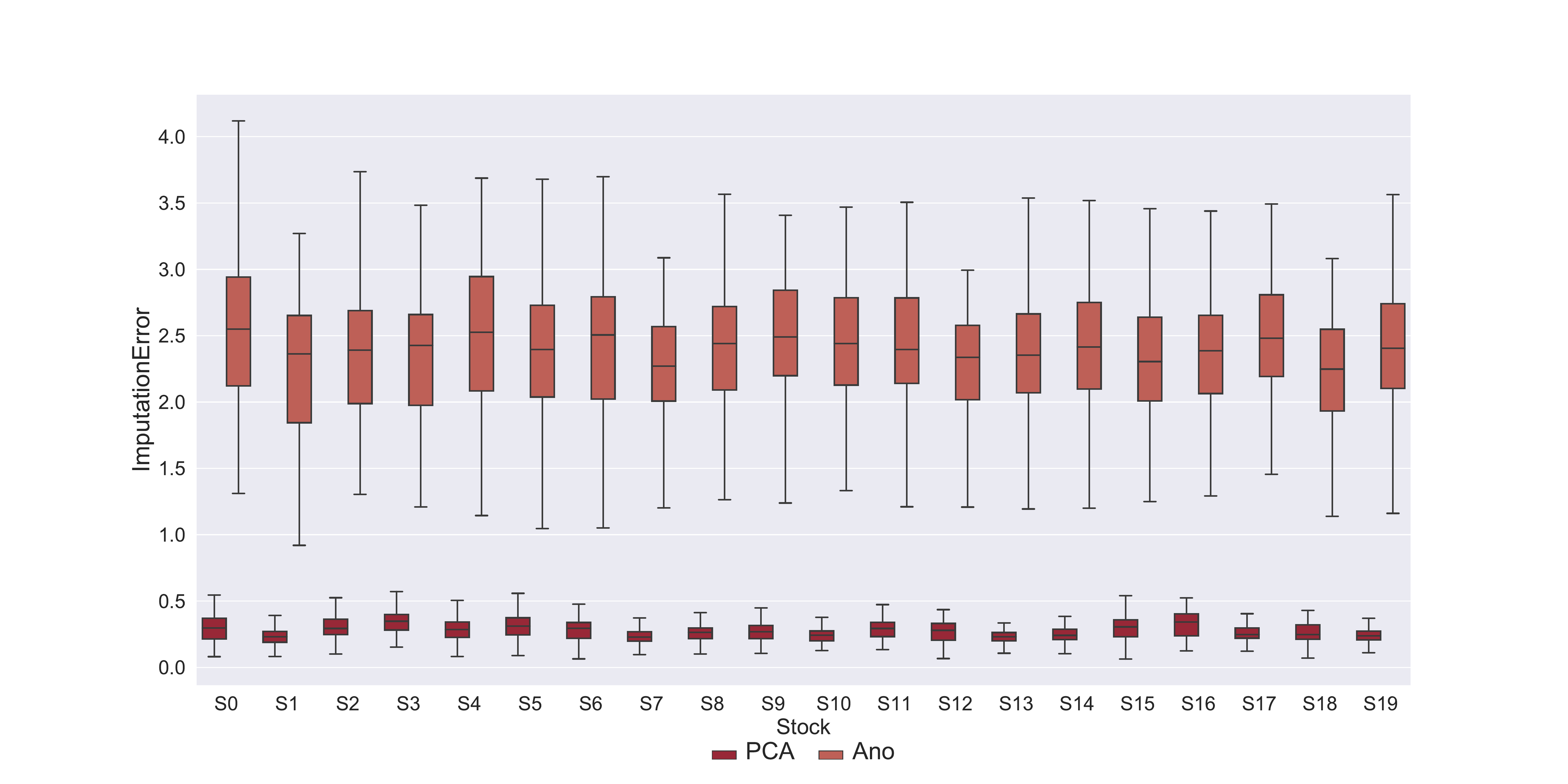}
    \caption{Boxplot representation of the distribution of the errors on time series imputation before the imputation of anomalies (Ano) and after replacing the anomalies with the reconstructed values suggested by PCA NN (PCA).}
    \label{fig:boxPCAvsNone}
\end{figure}

\begin{figure}
    \centering
    \includegraphics[width=\linewidth]{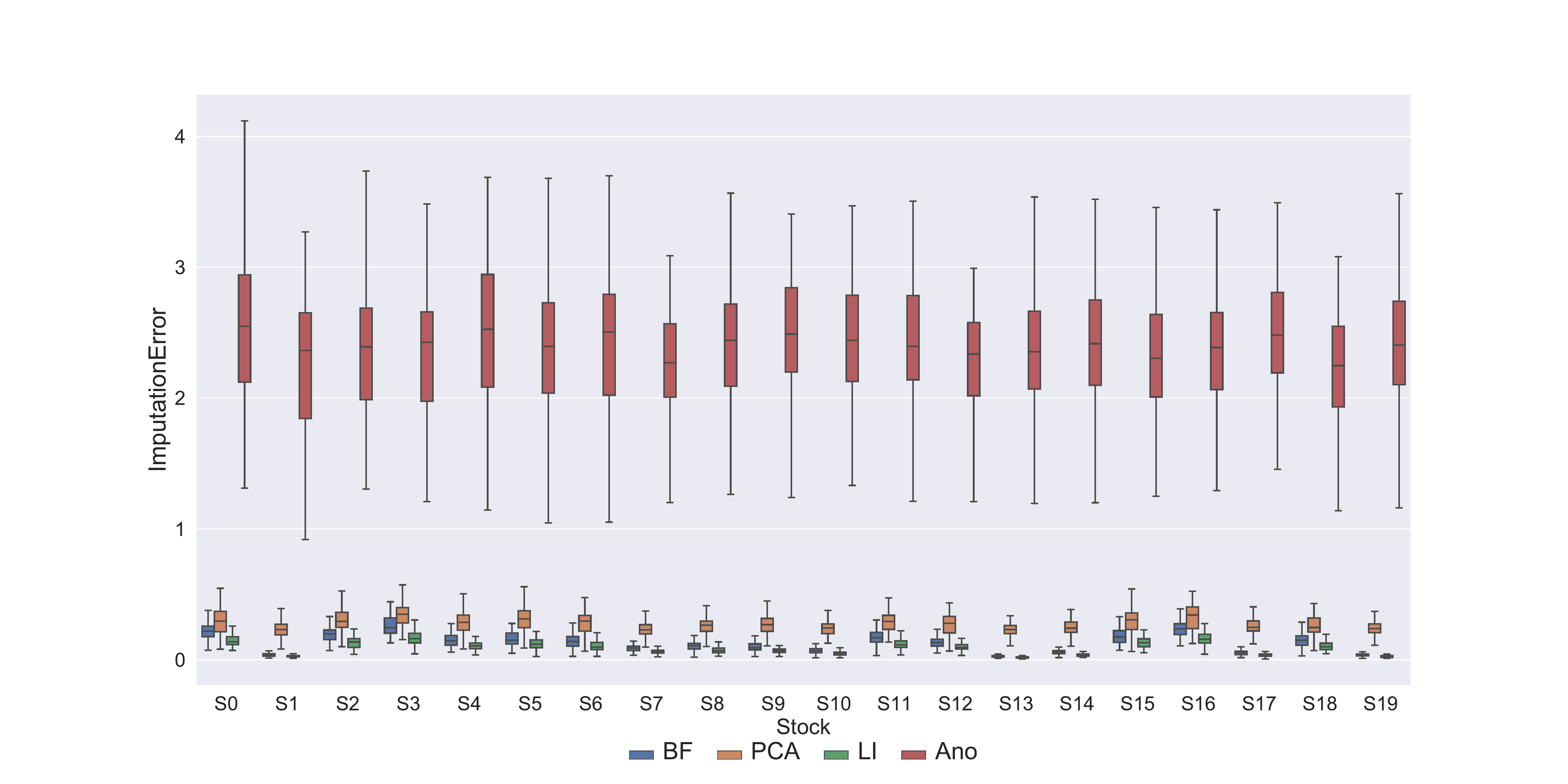}
    \caption{Boxplot representation of the distribution of the errors on time series imputation before the imputation of anomalies (Ano) and after replacing the anomalies; with the reconstructed values suggested by PCA (PCA), using linear interpolation (LI) and using backward fill (BF).}
    \label{fig:boxPCAvsAll}
\end{figure}

\begin{figure}
  \centering
  \begin{tabular}{cc}
  \includegraphics[scale=0.3]{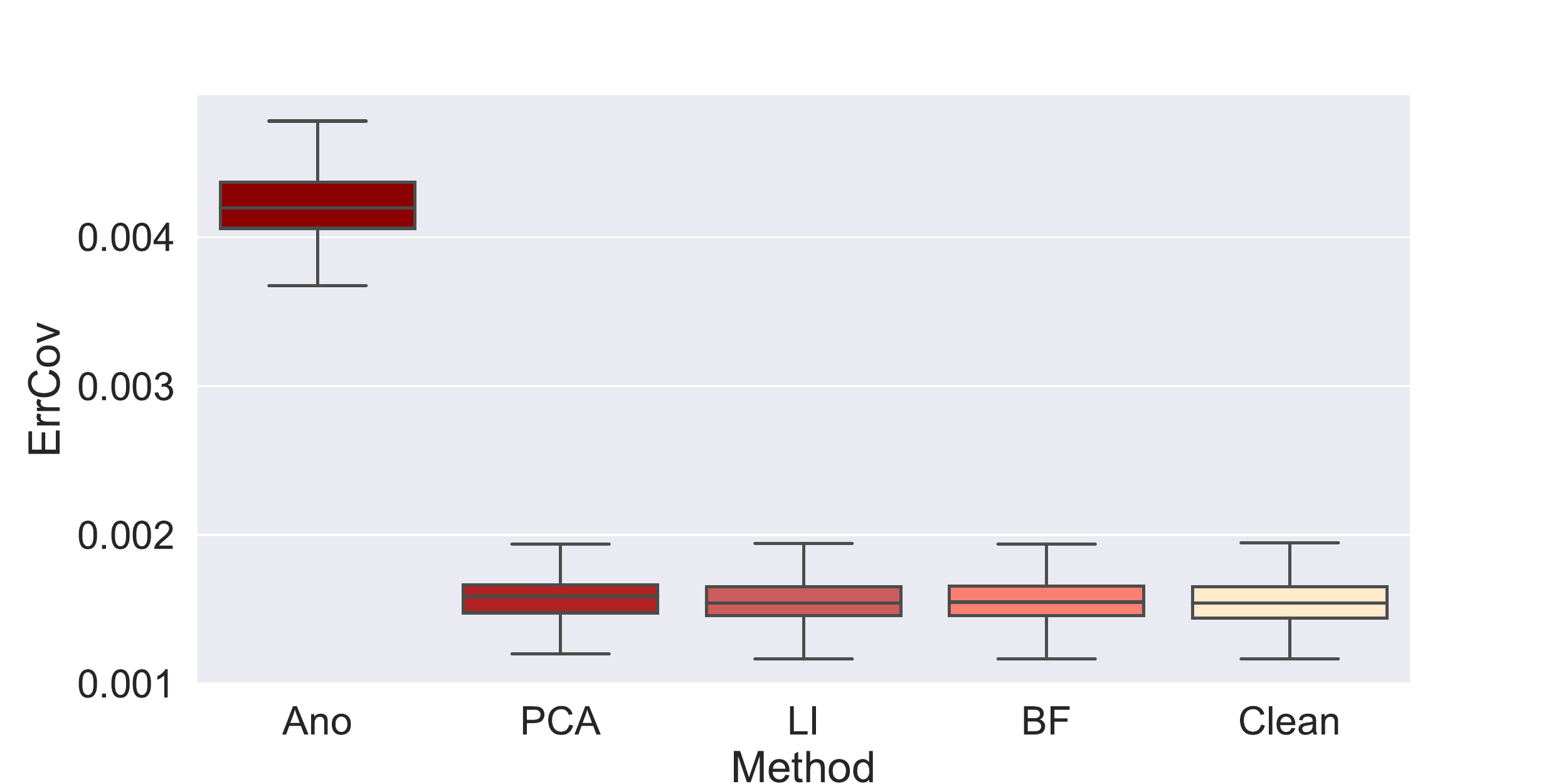} & \includegraphics[scale=0.3]{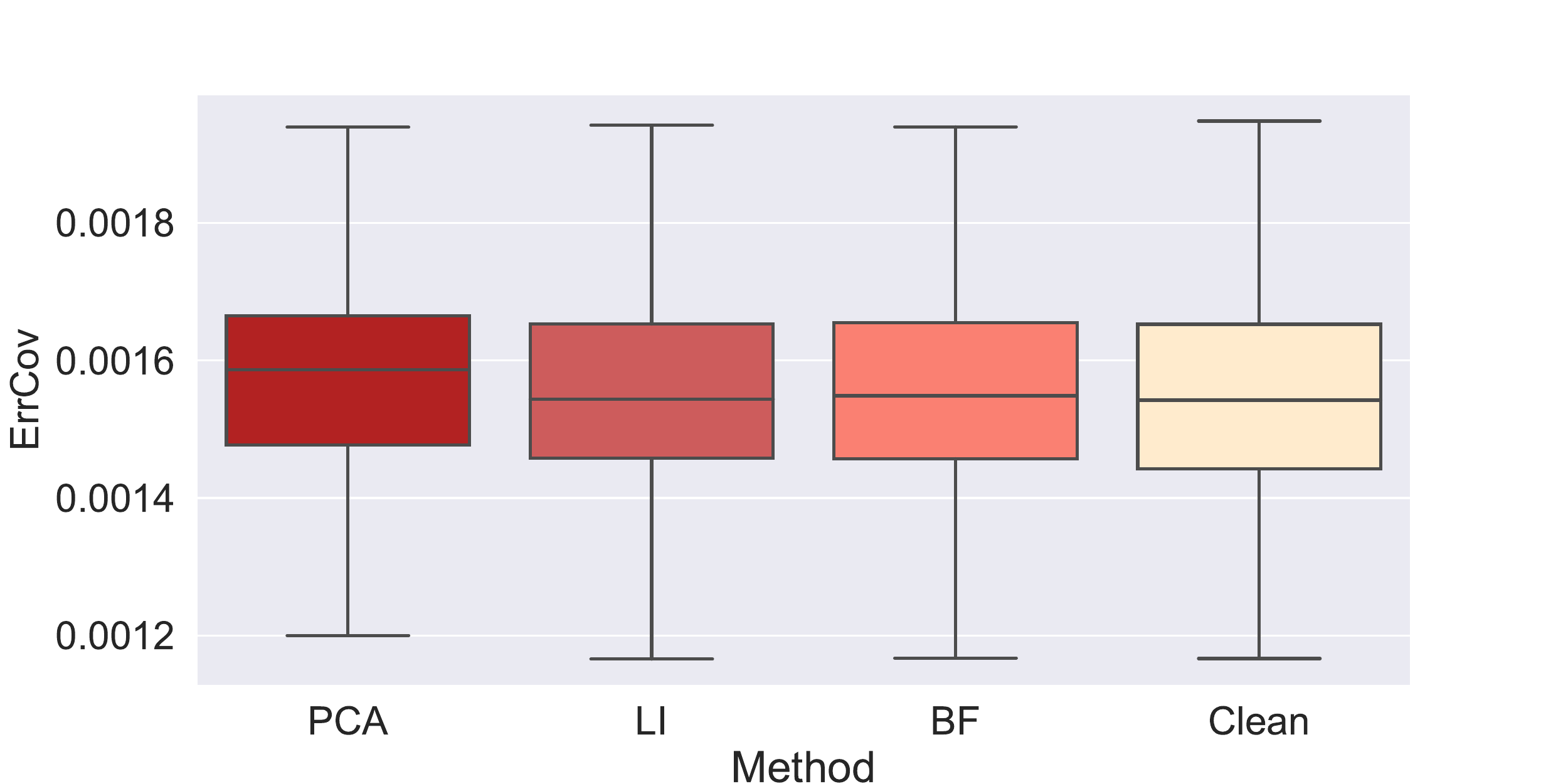} \\
  \end{tabular}
  \caption{Boxplot representation of the distribution of the errors on covariance matrix (on several stocks paths samples). Ano, PCA NN, LI, BF and Clean refer to the errors on the covariance when estimated on time series with anomalies, time series after  anomalies imputation following three approaches: imputation by the reconstructed values suggested by PCA (PCA),  using linear interpolation (LI), with backward fill (BF) and time series without anomalies (Clean). The Ano, PCA NN, LI, BF and Clean errors on the covariance estimation are all represented \textit{(left)}. For better visualization we remove the Ano errors \textit{(right)}.  } \label{fig:boximpcov}
\end{figure}

Each stock path was diffused and contaminated 100 times. The anomalies were then imputed following two baseline approaches (BF and LI). Figure \ref{fig:boxPCAvsNone} clearly shows that imputation using the reconstructed values with PCA reduces the imputation error but not as much as basic imputation techniques (LI and BF) as it is shown on Figure \ref{fig:boxPCAvsAll}. Table \ref{table:CovErrorImputation} shows that the baseline imputation approaches achieve even lower errors on the estimation of the covariance matrix. Figure \ref{fig:boximpcov} shows a higher mean but a lower variance of the errors for the PCA-based imputation approach. In conclusion, it would be rather recommended to replace the flagged anomalies with approaches such as backward fill or linear interpolation.

\begin{table}[H]
\centering
\begin{tabular}{c|ccccc}
Method & Ano  & PCA  & LI     & BF       & Clean         \\
\hline
Mean  & 0.004208&	0.001575&	0.001554&	0.001554&	0.001552\\
Standard deviation & 0.000272&	0.000165&	0.000170&	0.000172&	0.000173
\end{tabular}
\caption{Mean and standard deviation error on covariance matrix after imputation of anomalies with baseline methods.}
\label{table:CovErrorImputation}
\end{table}



\begin{figure}
    \centering
    \includegraphics[scale=0.5]{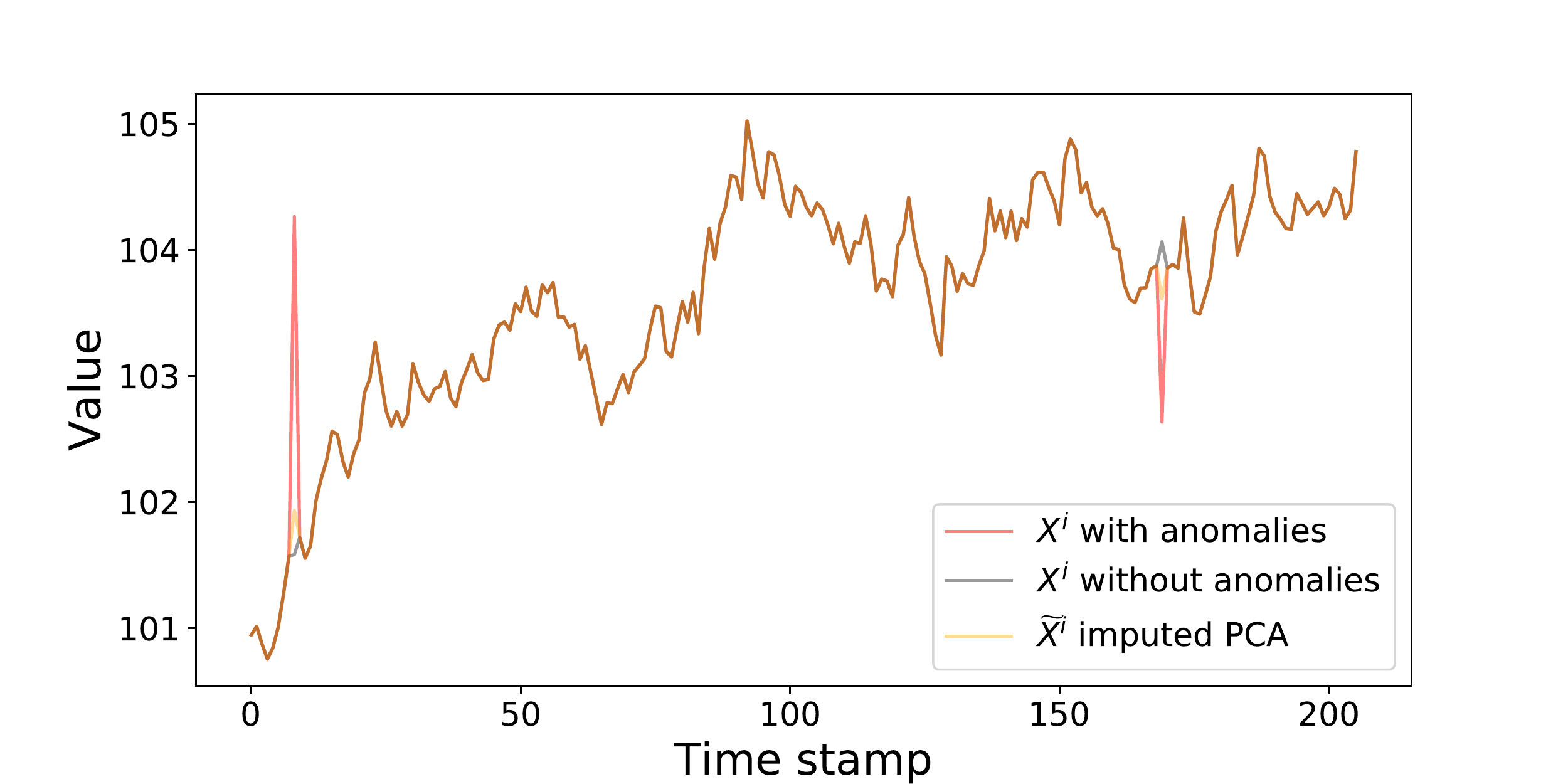}
    \caption{Original, contaminated and reconstructed stock price path.}
    \label{fig:ImpuPCAPath}
\end{figure}
\begin{figure}
    \centering
    \includegraphics[scale=0.6]{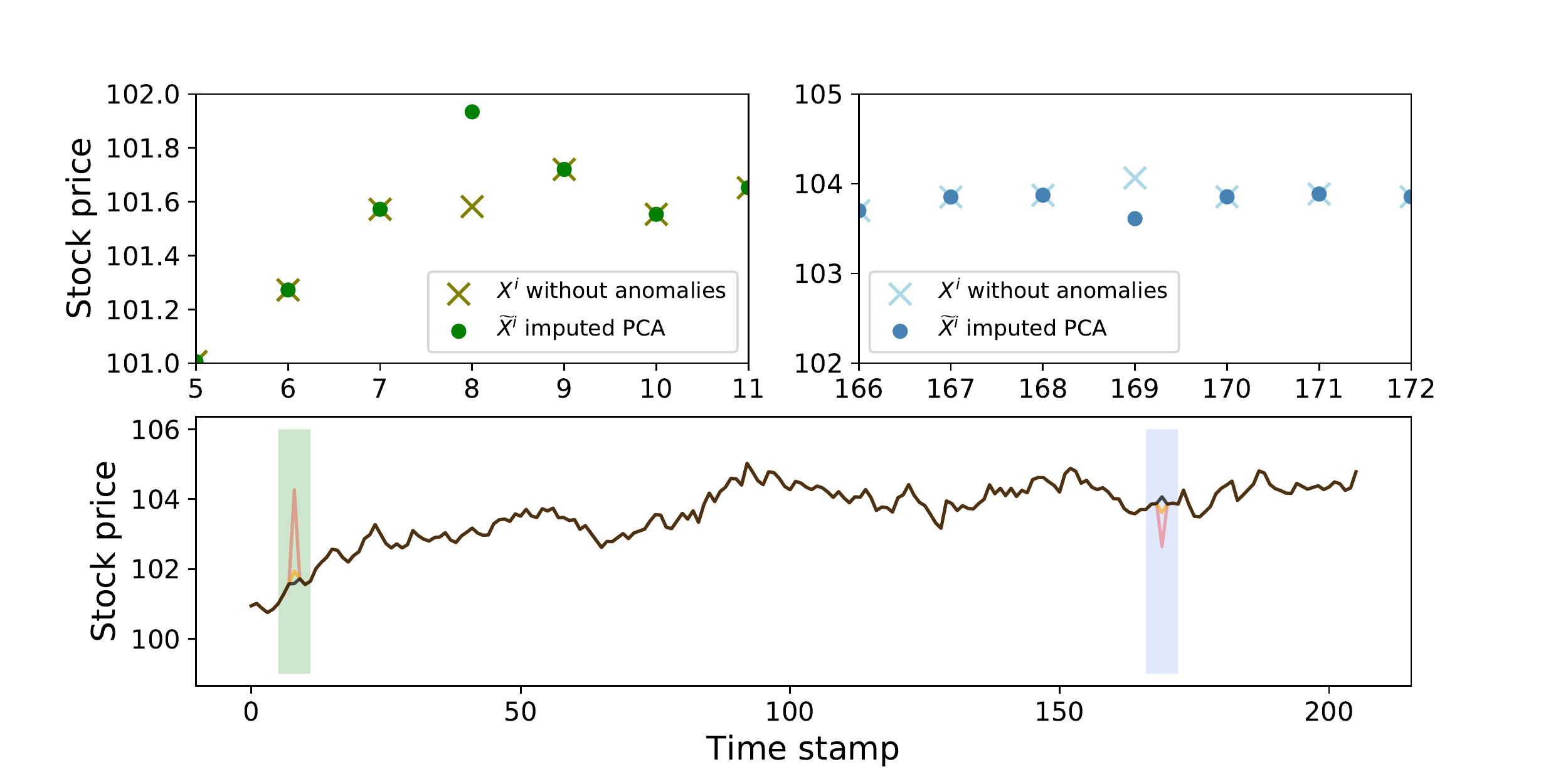}
    \caption{Original, contaminated and imputed stock prices path in black, red and orange. The two additional graphs represent the region around anomalies, where the cross and circle respectively show the true value of the stock and its value after imputation of the anomalies using the reconstructed value suggested by the PCA NN approach.}
    \label{fig:ImpuPCAPathZoomIn}
\end{figure}

A natural explanation to the relative inefficiency of the reconstruction value as imputation value comes from the fact that, by construction, the reconstructed value integrates in its computation the abnormal value, which is not the case when using other imputation techniques. Therefore, even if the reconstructed value is closer to the true value than it is to the abnormal value, the spread between the imputation and the true value is still significant. Figures \ref{fig:ImpuPCAPath} and \ref{fig:ImpuPCAPathZoomIn} illustrate the latter  with a plot of a reconstructed and original price path. 

\subsection{Cut-off Value Robustness}
\label{subsection:ThresholdRobustness}

In most anomaly detection models, the cut-off value is a hand-set parameter. In our approach,  the cut-off value is a model parameter and as such it is calibrated through the learning. Testing the robustness of the cut-off given by the model is therefore a must. The identification step is the unique step of the approach being concerned by the robustness, since it is the only step involving the cut-off calibration. Robustness is checked by shocking the suggested cut-off with different level of noise and observing the impact of these shocks on the model performance. We consider several shock amplitude $\gamma \in \pm \{10^{-4},10^{-3},10^{-2},10^{-1},1,2 \}$. Table \ref{tab:RobustCheckThreshold} reports the mean and standard deviation of the scores of the model. The cut-off calibrated on the synthetic training data sample is shocked. Scores on both train and test sets are computed using the shocked cut-off values. 

\begin{table}[H]
\centering
\begin{tabular}{c}
   \begin{tabular}{c|cccc}
$\gamma$ & Accuracy & Precision & Recall& $F_1$ \\
\hline 
$10^{-4}$         & 0.9097            & 0.9736             & 0.8421          & 0.9031            \\
-$10^{-4}$       & 0.9097            & 0.9736             & 0.8421          & 0.9031            \\
\hline
$10^{-3}$        & 0.9097            & 0.9738             & 0.8419          & 0.9031            \\
-$10^{-3}$        & 0.9098            & 0.9737             & 0.8424          & 0.9033            \\
\hline
$10^{-2}$        & 0.9092            & 0.9747             & 0.8403          & 0.9025            \\
-$10^{-2}$        & 0.9098            & 0.9720             & 0.8439          & 0.9035            \\
\hline
$10^{-1}$        & 0.9042            & 0.9860             & 0.8201          & 0.8954            \\
-$10^{-1}$        & 0.9103            & 0.9563             & 0.8599          & 0.9056            \\
\hline
0         & 0.9097            & 0.9736             & 0.8421          & 0.9031            \\
\hline
1         & 0.7771            & 0.9994             & 0.5546          & 0.7133            \\
-1        & 0.5183            & 0.5093             & 0.9998          & 0.6749            \\
\hline
2         & 0.6284            & 1.0000             & 0.2567          & 0.4086            \\
-2        & 0.5001            & 0.5000             & 1.0000          & 0.6667           \\
\end{tabular} 

\begin{tabular}{c|cccc}
 $\gamma$ & Accuracy & Precision & Recall & $F_1$ \\
\hline 
$10^{-4}$ & 0.8858            & 0.6126             & 0.8527          & 0.7130            \\
$-10^{-4}$ & 0.8858            & 0.6126             & 0.8527      & 0.7130    \\
\hline 
$10^{-3}$ &0.8858            & 0.6126             & 0.8527          & 0.7130            \\
$-10^{-3}$ &0.8850            & 0.6105             & 0.8527          & 0.7116            \\
\hline 
$10^{-2}$ &0.8877            & 0.6179             & 0.8527          & 0.7166            \\
$-10^{-2}$ & 0.8838            & 0.6074             & 0.8527          & 0.7095            \\
\hline 
$10^{-1}$ & 0.8984            & 0.6502             & 0.8432          & 0.7342            \\
$-10^{-1}$ & 0.8672            & 0.5653             & 0.8741          & 0.6866            \\
\hline 

0 & 0.8858            & 0.6126             & 0.8527          & 0.7130            \\
\hline 

1 & 0.9391            & 0.9435             & 0.6746          & 0.7867            \\
-1 & 0.2660            & 0.1848             & 1.0000          & 0.3120            \\
\hline 
2 & 0.8949            & 0.9814             & 0.3753          & 0.5430            \\
-2 & 0.1700            & 0.1670             & 1.0000          & 0.2862           \\
\end{tabular}\\
\end{tabular}
        \caption{Performance evaluation after shocking the calibrated cut-off value for the training set \textit{(left)} and for the test set \textit{(right)}.}
\label{tab:RobustCheckThreshold}
\end{table}

 Excluding the extreme cases where the shock amplitude is $\pm\{1,2\}$, one could see that the accuracy is barely impacted by the shocked cut-off values, both on the train and test sets. The interpretation of the remaining results is split in two and is applicable for both the training and test sets. 

 When  negative shocks are applied, the model predicts more anomalies and less normal observations (compared to the predicted numbers with the calibrated cut-off value). Therefore, the model is able to identify anomalies which were missed initially. Hence, applying negative shocks increases the recall. As for the precision, i.e. the rate of correctly identified contaminated time series, since on left hand side of the cut-off value we are in the density region where contaminated time series represent the dominant class, the new position of the cut-off value leads to a misidentification of this type of observations. This entails a deterioration of the precision. 
Opposite behaviours of the precision and recall are observed when positive shocks are applied, since some abnormal time series are missed by the model. 

\begin{figure}
  \centering
  \begin{tabular}{cc}
  \includegraphics[scale=0.3]{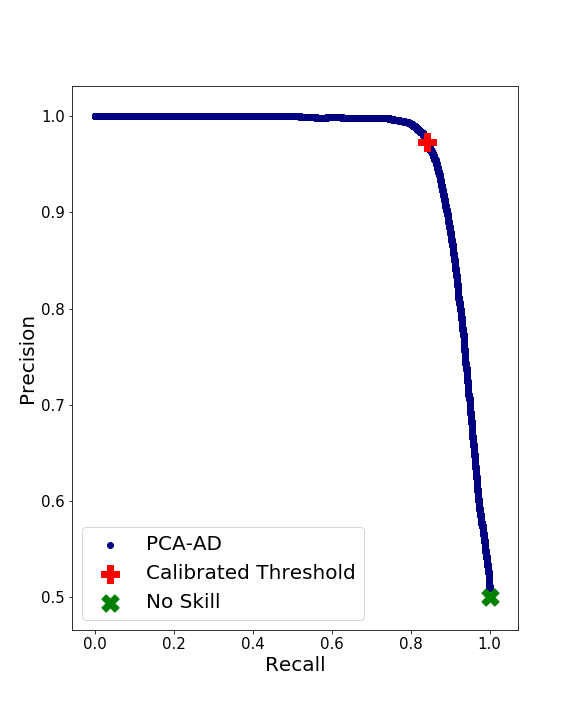}&
  \includegraphics[scale=0.3]{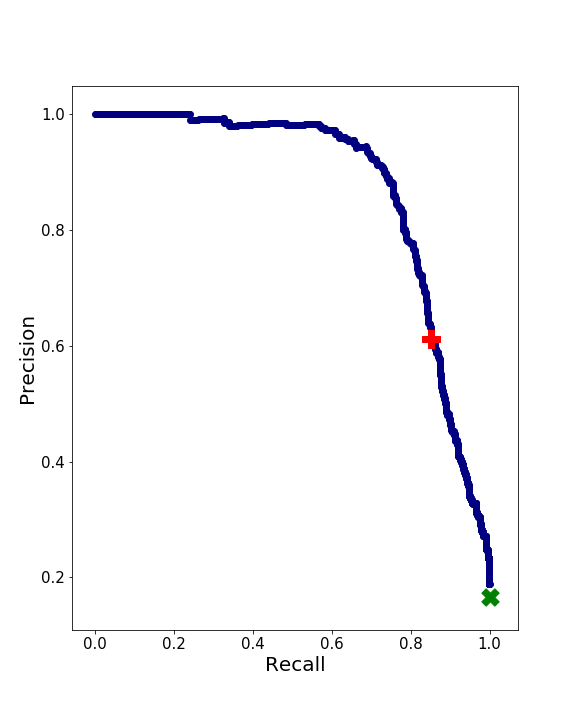} 
  \end{tabular}
  \caption{Precision-recall curve of the suggested model (in blue), scores with the calibrated threshold (red plus), no-skill model scores (green cross) for the training set \textit{(left)} and a test set \textit{(right)}.} \label{fig:precisionrecallcurve}
\end{figure}

Figure \ref{fig:precisionrecallcurve} shows the precision and recall when cut-off values other than the one we calibrated are used in our model, on the train and test sets. These scores are also compared with the performance of the no-skill model\footnote{The no-skill model assigns to all observations the positive label.}. We note that our approach edges out the no-skill model on both sets. The area under the curve (AUC) score\footnote{AUC score ranges from 0 to 1, 1 being the score associated with a perfect model.} on the train set and the test set are respectively  0.97 and 0.87. This shows that we are better performing on the train set, which was expected, but the performance on test set is just as satisfying. These results reinforce our conclusions regarding the robustness of our approach. One can see that the calibrated cut-off value represents, for the train set, the point where an equilibrium is being reached between precision and recall. The calibrated cut-off allows to achieve high precision and recall scores,  simultaneously. We conclude that the cut-off given by the model is suitable for the training samples and for unseen samples as well.

To briefly sum up this section, although we  mentioned that some shocks on the cut-off value induce higher scores, the improvement over the scores with the calibrated cut-off value is not significant (unless high amplitude shocks are considered). The numerical tests and the precision-recall curves are consistent with the robustness of the cut-off value suggested by the approach.

\subsection{Sensitivity to Anomaly Amplitude}

Time series are manually contaminated as described in Section \ref{subsec:DataContamination}. The abnormal value $S^{a,i}_{t_{\jmath}}$ of the $i$-th time series results from a shock of the initial value of the time series denoted by $S^i_{t_{\jmath}}$ as stated in \eqref{eq:contam}. We recall that the shock is represented by $\delta$, whereas its amplitude $\lvert \delta \rvert$ is uniformly drawn from $[0,\rho]$. Hence, $\rho$ is the parameter that ultimately controls the amplitude of the anomaly. Since $\rho$ is fixed by the user, it is interesting to investigate the sensitivity of the PCA NN approach performance to the amplitude of the shock.


The PCA NN approach evaluated in Sections \ref{subsection:evalstep1}-\ref{subsection:evalstep2} is calibrated on time series which were contaminated with shock amplitude drawn from $\mathcal{U}\left([0,\rho]\right)$ with $\rho=0.04$. For the data sets on which the model was calibrated and then evaluated, the anomalies were grouped according to the shock amplitude they result from. For the identification step, we distinguish four groups of contaminated times. For instance, the first line of Table \ref{tab:sensiAmpliID} defines the first group of contaminated time series, for which the anomalies results from shock amplitude $\in {[}0.3091e-2, 1.46e-2{[}$. For this first group with this specific range of amplitude, $77\%$ of the contaminated time series were identified during the PCA NN identification step.

\begin{table}[H]
\centering
\begin{tabular}{c|c}
Amplitude Range ($10^{-2}$)  & Detection ratio \\
\hline
{[}0.309, 1.46{[} & 0.77       \\
{[}1.46, 2.34{[} & 0.91       \\
{[}2.34, 2.92{[} & 0.96       \\
{[}2.92, 3.78{]} & 0.98      
\end{tabular}
\caption{Detection ratio of correctly identified contaminated time series on test set, with time series being grouped according to their anomalies shock amplitude.}
\label{tab:sensiAmpliID}
\end{table}

Similarly, for the localization step, we consider four groups of anomalies. The first line of Table \ref{tab:sensiAmpliLoc} represents the anomalies with shock amplitude $\in {[}0.309e-2, 1.31e-2{[}$. $86\%$ of the anomalies belonging to this first group were correctly localized by the PCA NN localization step. 

\begin{table}[H]
\centering
\begin{tabular}{c|c}
Amplitude Range ($10^{-2}$)   & Detection ratio  \\
\hline 
{[}0.309, 1.31{[} & 0.86        \\
{[}1.31, 2.29{[} & 1.00         \\
{[}2.29, 2.88{[} & 1.00         \\
{[}2.88, 3.78{[} & 1.00       
\end{tabular}
    \caption{Detection ratio of correctly localized anomalies on test set, with anomalies being grouped according to their shock amplitude.}
    \label{tab:sensiAmpliLoc}
\end{table}

The bounds defining each group are chosen to be the minimal value, $25\%$-quantile, $50\%$-quantile, $75\%$-quantile, and the maximum value over the shock amplitude. From the results of Tables \ref{tab:sensiAmpliID} and \ref{tab:sensiAmpliLoc}, one could see that, except for the first group which represents the lowest shock amplitude, the detection ratio of the remaining groups are similar.  

\begin{figure}[H]
    \centering
    \includegraphics[scale=0.3]{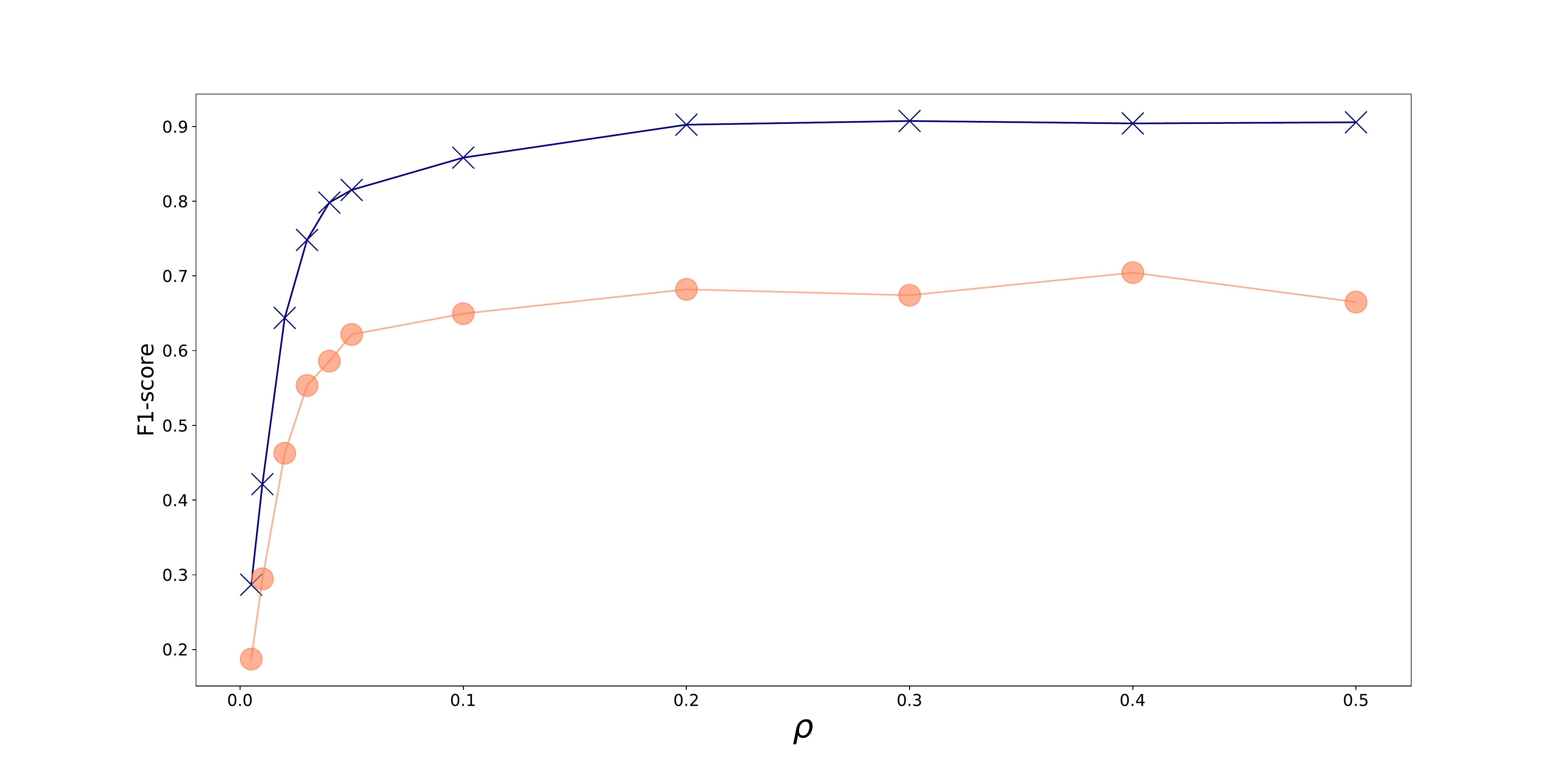}
    \caption{Mean $F_1$-score for identification step with respect to various values of $\rho$ on train and test set, respectively represented  by the blue and pink curves.}
    \label{fig:sensiAmplID}
\end{figure}

\begin{figure}
    \centering
    \includegraphics[scale=0.3]{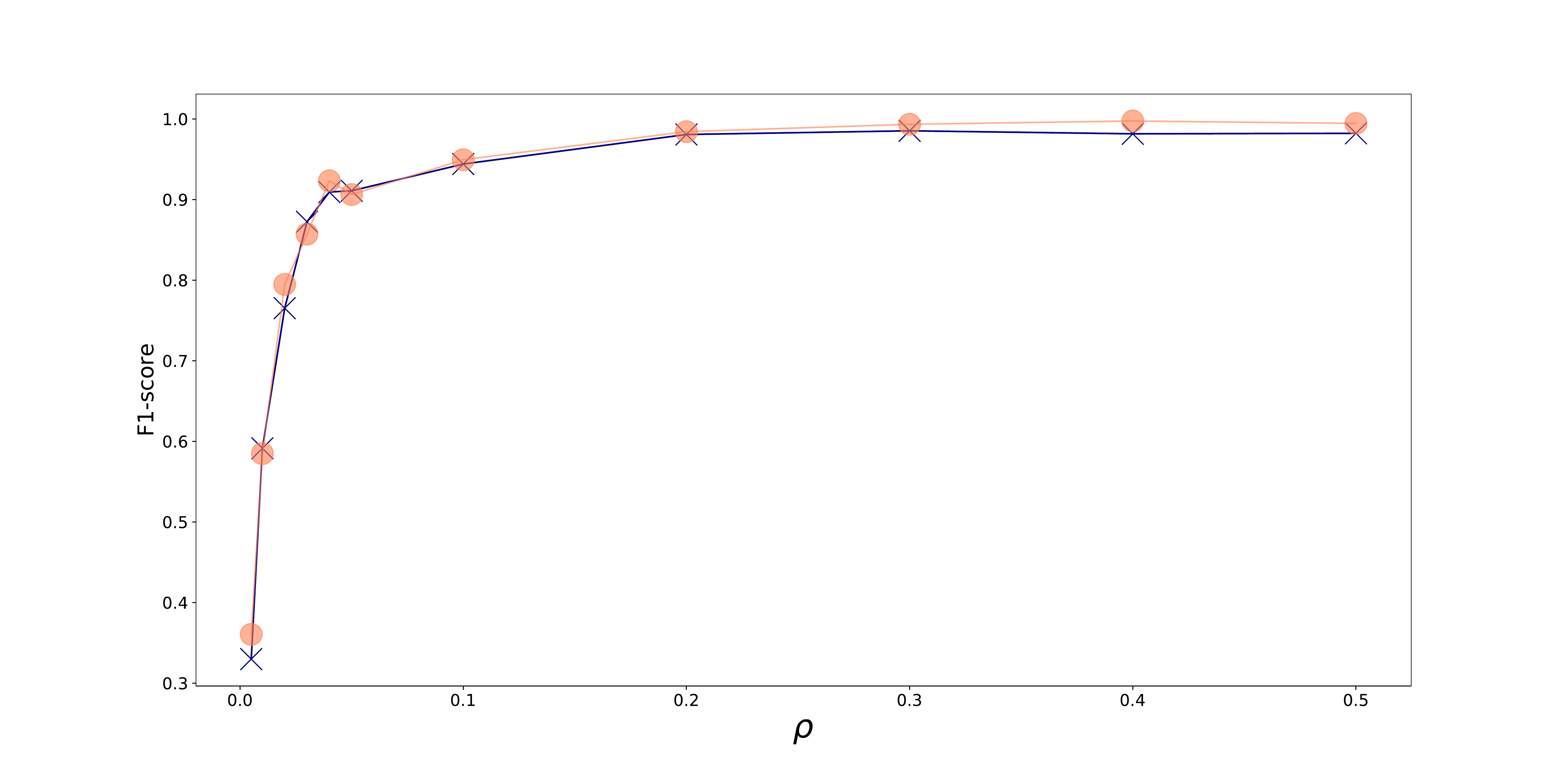}
    \caption{Mean $F_1$-score for localization step with respect to various values of $\rho$ on train and test set, respectively represented  by the blue and pink curves.}
    \label{fig:sensiAmplLoc}
\end{figure}

The calibrated PCA NN approach performance is then tested on new data sets contaminated with distinct values of $\rho$. Figures \ref{fig:sensiAmplID} and \ref{fig:sensiAmplLoc} show that the $F_1$ scores are almost similar across the high values of $\rho$, both for the train and test sets. Moreover, Figure \ref{fig:sensiAmplLoc} clearly states similar and high performance of the localization stage on both data sets, regardless of the value of $\rho$.

To conclude on the sensitivity of the model performance to the amplitude of the anomaly shocks, although it is clear that a lower performance is observed on the identification and localization of smaller amplitude shock anomalies, the detection ratio is still satisfactory for this category of anomalies. 

\section{Application to a  Downstream Task: Value-at-Risk Computations }
\label{section:VaR}

\def\ell{\varrho} 

In this section, we illustrate the benefit drawn when applying the PCA NN approach as a pre-processing before value-at-risk computations.  

Given a random variable $\ell$  representing the loss in portfolio position over a time horizon $h$, its value-at-risk at the confidence level  $\alpha \in (\frac{1}{2},1)$, $\mathrm{VaR} _{\alpha}
(\ell)$,
is defined by
the quantile of level $\alpha$ of the loss distribution, i.e. $\mathbb{P}(\ell \le \mathrm{VaR} _{\alpha}
(\ell))=\alpha$ (assuming $\ell$ atomless for simplicity).
Let $(S_t)_{t=t_1, \ldots, t_T}$ be a path price diffusion distributed according to the Black-Scholes model \eqref{eq:diffeq}. The logarithmic returns  to maturity are distributed according to the Gaussian distribution

\begin{align}
    \ln\left(\frac{S_{t+h}}{S_t}\right) \sim \mathcal{N} \left( \left(\mu - \frac{\sigma^2}{2}\right)h, \sigma^2h\right),
\end{align}
with $\mu \in \mathbb R$ and $\sigma >0$.




We assume  the vector $\bm{R}$ of log-returns on our stocks to be joint-normal,
\[
\bm{R} = 
\left(
R^{1}, R^{2},  \ldots , R^{i}, \ldots, R^N
\right)^{\top}  \sim \mathcal{N}_N\left(\mu_R,\Sigma_R\right),
\]
where $R^i = \log\left(\frac{S^i_{t+h}}{S^i_t}\right) \sim \mathcal{N}\left(\mu_{i,R},\sigma_{i,R}^2\right)$ and $\Sigma_R$ is the covariance matrix. 

We consider a portfolio on $N$ stocks, which return is given by $P = \mathcal{Q}^\top \bm{R}$, where $\mathcal{Q} \in \mathbb{R}^N$ defines the composition of the portfolio. Hence, $P \sim \mathcal{N}\left(\mu_P ,\sigma_P^2\right)$, where $\mu_P =\mathcal{Q}^\top \mu_R$ and $\sigma_P^2 = \mathcal{Q}^\top  \Sigma_R \mathcal{Q}$. 
The value-at-risk for the time horizon $h$ at level $\alpha$ is

\begin{align}
    \mathrm{VaR}_{\alpha}(P)
 = \mu_P +  q_{\alpha}  \sigma_P \; ,
    \label{eq:VaR}
\end{align}
where $q_{\alpha}$ is the $\alpha$-quantile of a standard normal distribution and the parameters $ \mu_R$ and $ \Sigma_R$ are estimated from different types of time series to evaluate the impact of localizing and removing anomalies following our approach.

Under the adopted framework, the true $\mathrm{VaR}_{\alpha}(P)$, $\mathrm{VaR}^{theo}$ is thus known and can be computed using the diffusion parameters. An estimation $\widehat{\mathrm{VaR}}_{\alpha}(P)$ can be obtained by replacing the parameters in \eqref{eq:VaR} by their estimates $\widehat{\mu_P}$ and $\widehat{\sigma_P}$ computed from the time series with anomalies, or after imputation of anomalies. Then, absolute errors 
and relative errors are computed as 

\begin{align*}
  &  \text{AbsoluteError}_{\text{VaR}} = \left \lvert \mathrm{VaR}_{\alpha}(P) - \widehat{\mathrm{VaR}}_{\alpha}(P)\right \rvert,\\
&
    \text{RelativeError}_{\text{VaR}} = \frac{\left \lvert \mathrm{VaR}_{\alpha}(P) - \widehat{\mathrm{VaR}}_{\alpha}(P)\right \rvert}{\mathrm{VaR}_{\alpha}(P)}
\end{align*}

Table \ref{tab:VaRnames} summarizes the four VaR estimations we consider in the sequel:

\begin{table}[H]
\begin{tabular}{c|l}
VaR estimation Name       & $\widehat{\mu_R}, \widehat{\Sigma_R}$ estimated on \\ \hline
$\mathrm{VaR}^{clean}$    & Time series without anomalies                                                                                      \\ 
$\mathrm{VaR}^{anom}$      & Times series with anomalies                                                                                        \\ 
$\mathrm{VaR}^{loc,true}$ & Time series after anomalies imputation knowing their true localization                                             \\ 
$\mathrm{VaR}^{loc,pred}$ & Time series after anomalies imputation based on predicted localization                                             \\ 
\end{tabular}
\caption{Notations of VaR estimation based on the time series the VaR parameters have been estimated from.}
\label{tab:VaRnames}
\end{table}


To conduct this analysis, we generate new stock path samples that we assume to be clean, fixing the diffusion parameters $\mu_R$ and $\Sigma_R$. We then add the anomalies following the procedure described in Section \ref{section: data}. We apply our model to localize the anomalies and replace the localized anomalies using the backward fill (BF) approach (shown to be the more efficient imputation technique in Section \ref{subsection:imputation}). For each stock and each run of simulation, we obtain four estimates of the distribution parameters of the associated log-returns.

\begin{figure}
    \centering
    \includegraphics[scale=0.3]{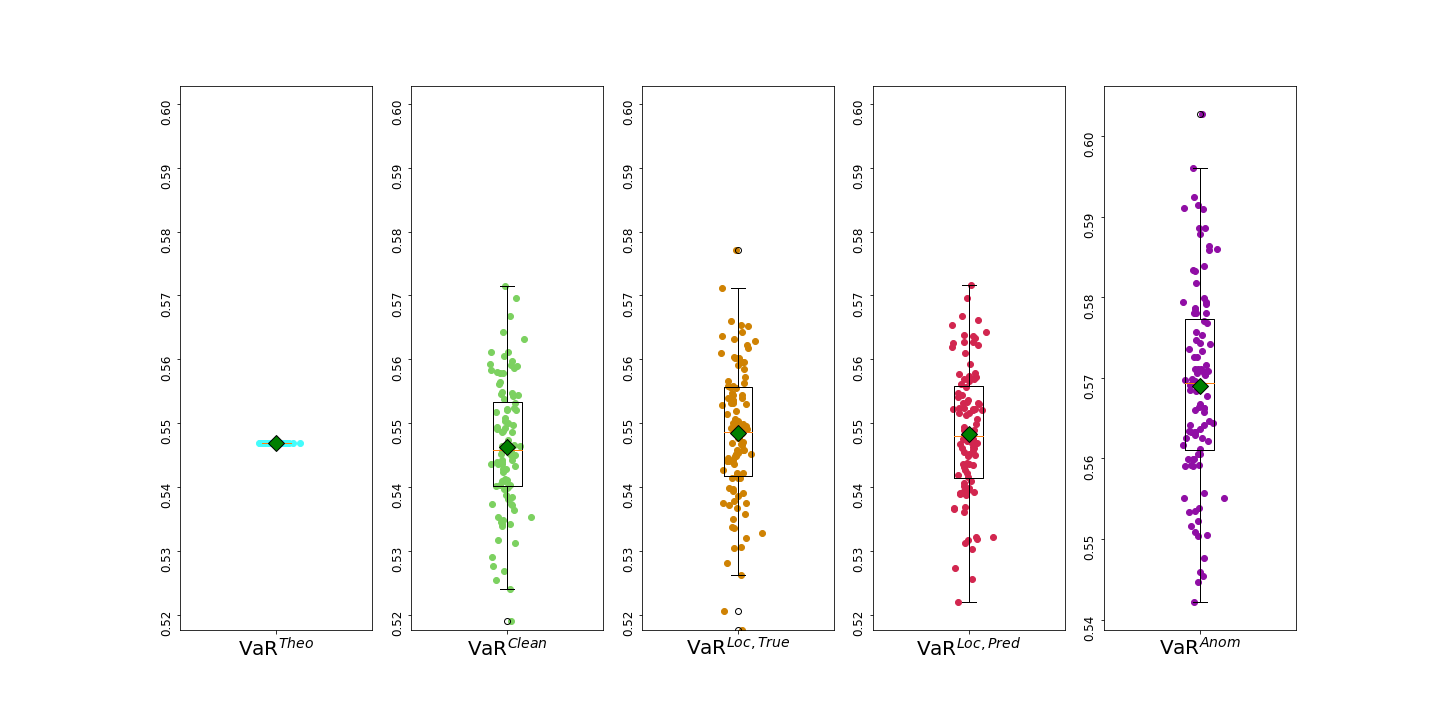}
    \caption{Boxplot representation of parametric VaR estimations for $P$. The green square represents the mean of VaR estimation.}
    \label{fig:VaRonPortfolioBox}
\end{figure}

\begin{table}[H]
\begin{tabular}{c|c|c|c|c|c}
VaR               & $\mathrm{VaR}^{theo}$ & $\mathrm{VaR}^{clean}$ &$\mathrm{VaR}^{loc,true}$ &$\mathrm{VaR}^{loc,pred}$&$\mathrm{VaR}^{anom}$ \\ \hline
Mean               & 	0.546851&	0.546300&	0.548392&	0.548270&	0.569015\\ 
Standard Deviation &0.0&	0.010105&	0.010739&	0.010832 &	0.012268          \\ 
\end{tabular}
\caption{Summary of VaR estimations for $P$.}
\label{table:VaRonPortfolioBox}
\end{table}

Figure \ref{fig:VaRonPortfolioBox} and Table \ref{table:VaRonPortfolioBox} summarize the distribution of the VaR estimates for $\alpha = 0.99$ and $h= 1(day)$, over several simulation runs. The boxplots show the dispersion of the portfolio VaR estimates on several diffusions. The green square represents the mean of the VaR estimates. For the four first boxplots, the means are approximately on the same level, which is confirmed by the results of Table \ref{table:VaRonPortfolioBox}. The anomalies present among the time series observed values have a non-negligible impact on the distribution parameters estimation, which ultimately causes a wrong estimation of the VaR. Thanks to the localization of the anomalies by the suggested model and their imputation as per Section \ref{subsection:imputation}, we are able to get a more accurate estimation of the VaR. $\mathrm{VaR}^{loc,true}$ and $\mathrm{VaR}^{loc,pred}$ are quite similar, which shows that the model accurately localizes the anomalies. 
\begin{table}[H]
\centering
\begin{tabular}{c|c|c|c|c}
VaR & $\mathrm{VaR}^{clean}$ &$\mathrm{VaR}^{loc,true}$ &$\mathrm{VaR}^{loc,pred}$ &$\mathrm{VaR}^{anom}$\\ 
\hline 
Absolute Error & 	0.007995&	0.008596&	0.008622  &	0.02235     \\ 
Relative Error & 0.014620&	0.015720&	0.015767 &	0.04087            \\
\end{tabular}
\caption{Mean absolute and relative error on VaR estimations for $P$.}
\label{table:MAEMREVaRPortfolio}
\end{table}

We also evaluate the error on the VaR estimation using the mean absolute error and the mean relative error, taking the $\mathrm{VaR}^{theo}$ as our benchmark. As one can tell from Table \ref{table:MAEMREVaRPortfolio}, even when the distribution parameters are estimated from the clean time series, the VaR computed with these parameters is not exactly the one computed with the theoretical parameters. This can be explained by the historical size of the observed values used to estimate the parameters. This table shows that by removing anomalies we can reduce by a factor two the error on VaR estimation.

\begin{figure}
    \centering
    \includegraphics[scale=0.35]{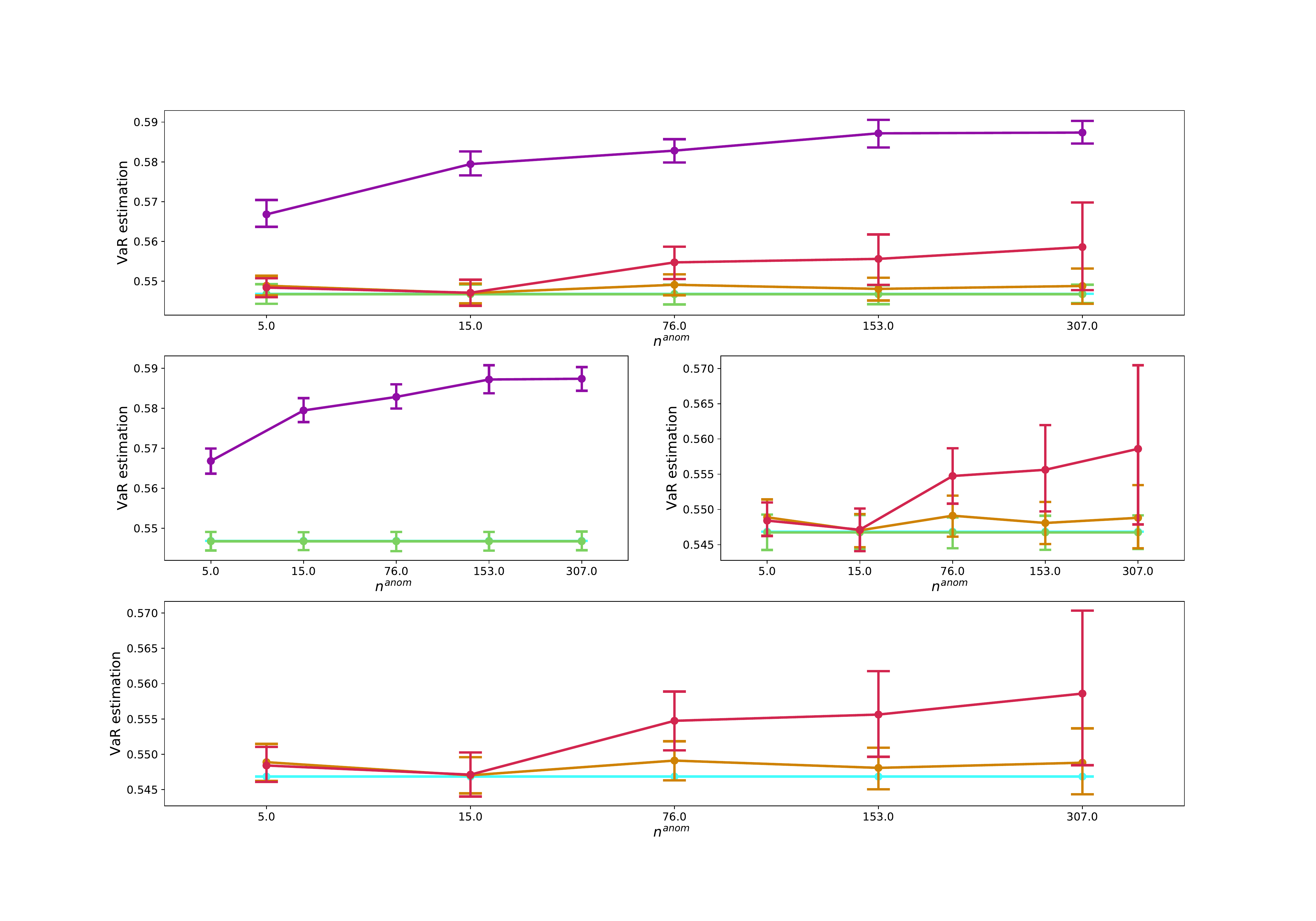}
    \caption{VaR estimation based on parameter estimation from time series with and without anomalies (purple and light blue curves), time series after localization and imputation of anomalies with the suggested approach (red curve) and time series imputed knowing the true localization of the anomalies (brown curve), with respect to $n^{anom}$.}
    \label{fig:boxVaRwrtContamRate}
\end{figure}

Additionally, we assess the impact on  $\mathrm{VaR}^{theo},\mathrm{VaR}^{clean},\mathrm{VaR}^{anom},\mathrm{VaR}^{loc,true}$ and $\mathrm{VaR}^{loc,pred}$ of increasing $n^{anom}$. To this end, we perform 50 simulation runs of stock paths for $n^{anom}$ and for each of those scenarios we estimate the VaR on the portfolio. We summarize the results on Figure \ref{fig:boxVaRwrtContamRate}, where each curve represents the mean VaR estimation with respect to $n^{anom}$ along with a representation of the uncertainty around each evaluated point through a confidence interval. 

When we compute the VaR using the abnormal time series,  we notice that the difference between $\mathrm{VaR}^{anom}$  and $\mathrm{VaR}^{theo}$ increases with $n^{anom}$, which is natural to expect. However, when the time series are cleaned prior to VaR estimation, the curve representing the VaR estimates are much closer to the ones representing $\mathrm{VaR}^{theo}$ and $\mathrm{VaR}^{clean}$, showing an undeniable improvement in the accuracy of the VaR estimation over the estimation based on abnormal time series. Furthermore, VaR estimation after the imputation following the model prediction or knowing the true localization of the anomalies seem to be quite similar for low $n^{anom}$, while some discrepancies between the two become more significant as $n^{anom}$ increases. A natural explanation could be that when the number of anomalies increases and the model suggests wrong anomalies localization, normal values are being replaced while true anomalies remain among the observed values, which wrongly impacts the VaR estimation. However, the results of Tables \ref{tab:meanRErrorVaR} and \ref{tab:stdRErrorVaR} indicate that the anomaly localizations suggested by the model are overall correct and allow removing most of the anomalies, as the relative error of $\mathrm{VaR}^{loc,pred}$ is, regardless of $n^{anom}$, always lower than the relative error of $\mathrm{VaR}^{anom}$ (e.g a relative error of 0.0248 for $\mathrm{VaR}^{loc,pred}$, against 0.0658 for $\mathrm{VaR}^{anom}$, when there are $76$ anomalies among the $1,500$ observed values of the time series).

\begin{table}[H]
\centering
\begin{tabular}{c|cccc}
$n^{anom}$ &$\mathrm{VaR}^{clean}$ &$\mathrm{VaR}^{loc,true}$ &$\mathrm{VaR}^{loc,pred}$&$\mathrm{VaR}^{anom}$ \\
\hline 
5       & 0.012194		&0.013796	&0.013341&0.037392        \\
15      & 0.012194		&0.012623	&0.016676  &0.059604      \\
76    & 0.012194		&0.014831	&0.024807   &0.065791     \\
153     & 0.012194		&0.014644	&0.034361   &0.073750     \\
307     & 0.012194		&0.020537	&0.052244     &0.074091  \\
\end{tabular}
\caption{Mean relative error of parametric VaR estimations with respect to $n^{anom}$, for parameters estimated from  clean time series, time series with anomalies, imputed time series following predicted location ($\mathrm{VaR}^{loc,pred}$) and true anomalies localization ($\mathrm{VaR}^{loc,true}$).  }
\label{tab:meanRErrorVaR}
\end{table}

\begin{table}[H]
\centering
\begin{tabular}{c|cccc}
$n^{anom}$ & $\mathrm{VaR}^{clean}$ &$\mathrm{VaR}^{loc,true}$ &$\mathrm{VaR}^{loc,pred}$ &$\mathrm{VaR}^{anom}$ \\
\hline
5       & 0.010356		&0.011511	&0.009678  &0.020674      \\
15      & 0.010356		&0.010514	&0.012723 &0.020172      \\ 
76    & 0.010356		&0.011842	&0.018696 &0.019930       \\
153     &0.010356		&0.013683	&0.026117 &0.024243      \\
307     & 0.010356		&0.022526	&0.057903   &0.019207     \\
\end{tabular}
\caption{Standard deviation of relative error of parametric VaR estimations with respect to $n^{anom}$, for parameters estimated from  clean time series, time series with anomalies, imputed time series following predicted location ($\mathrm{VaR}^{loc,pred}$) and true anomalies localization ($\mathrm{VaR}^{loc,true}$).  }
\label{tab:stdRErrorVaR}
\end{table}

\section{Numerical Results on Real Data}\label{s:real}

We consider a labelled real data set including stock prices, bonds yields, CDS spreads, FX rates, and volatilities. These data were collected from a financial data provider for the period between 2018 and 2020. These data sets were labelled by experts. They are provided on  \url{https://github.com/MadharNisrine/PCANN} in the form of 132,000 time series (after augmentation). The train set is balanced, while the test set is imbalanced with 20\% of contaminated time series. 

To ensure the independence of the train and test data sets, we calibrate the PCA NN approach considering the time series of 2018 and 2019, while the 2020 time series are dedicated to model evaluation. 
The performance evaluation of the PCA NN approach after its calibration on real data sets is shown in Table \ref{tab:ResCaliReal}.
As visible from the upper panel, the PCA NN  performs quite well for the identification step. Once the contaminated time series are identified, the model is able to localize the anomaly with high accuracy, as reflected by the scores on the test set in the lower panel.

\begin{table}[H]

\centering
\begin{tabular}{c}
\begin{tabular}{l|l|l|l|l}

Data set & Accuracy &Precision & Recall  &$F_1$-score \\
\hline
Train set & 92.88 \% & 99.17\%   & 86.49\% & 92.40\% \\
Test set  & 88.15\%  & 72.45\%   & 46.10\% & 56.35\% \\
\end{tabular} \\
 \\~\\

\begin{tabular}{l|l|l|l|l}

Data set & Accuracy &Precision & Recall  &$F_1$-score \\
\hline
Train set & 99.48 \% & 99.49\%   & 99.48\% & 99.48\% \\
Test set  & 96.05\%  & 96.21\%   & 96.05\% & 95.92\% \\
\end{tabular}\\
\end{tabular}
\caption{Performance of the PCA NN identification step \textit{(upper part)} and localization step \textit{(lower part)} on the real data set.}

\label{tab:ResCaliReal}
\end{table}

\subsection{PCA NN against State of the Art Models  on Real Data}

Even if the $F_1$-score on the identification step is not that high, the results show that the related performance remains better than the one of alternative state of the art approaches. 

Tables \ref{tab:perfmethodsIdentificationReal} and \ref{tab:perfmethodsLocalizationReal}
compare the performance of the PCA NN approach against the state of the art models of Section \ref{apx:AppendixA} on our real data set. PCA NN outperforms the benchmark models in both steps, with overwhelming results for the anomaly localization step. 
\begin{table}[H]
\centering
\begin{tabular}{l|cccc}
Model& Accuracy  &Precision& Recall &$F_1$-score             \\
\hline
IF          &73.17\% & 18.12\%  & 17.53\% & 17.82\% \\
LOF            &82.33\% & 44.57\% & 26.62\%  &  	33.33\% \\
DBSCAN         & 22.74\% & 15.71\% & 83.77\% & 26.46\% \\
sig-IF &73.34\%& 18.15\%& 17.47\% & 17.80\% \\\hline
KNN          &70.26\% & 23.48\% &35.06\% &28.13\% \\
SVM        & 50.11\% & 24.46\% & 96.10\% &39.00\% \\

PCA NN     &88.15\%  & 72.45\%   & 46.10\% & 56.35\%   
\end{tabular}
\caption{Performance evaluation of unsupervised \textit{(upper part)} and supervised \textit{(lower part)} models for contaminated time series identification step. }
\label{tab:perfmethodsIdentificationReal}
\end{table}

\begin{table}[H]
\centering
\begin{tabular}{l|cccc}
Model& Accuracy  &Precision& Recall &$F_1$-score             \\
\hline
IF          &78.50\% & 2.173\%  & 98.35\% & 4.252\% \\
LOF            &87.04\% & 0.3713\% & 9.616\%  & 0.7151\% \\
DBSCAN         & N/A& N/A & N/A & N/A \\
sig-IF &N/A& N/A& N/A & N/A \\\hline
KNN          &99.85\% & 78.38\% &95.59\% &86.13\% \\
SVM        & N/A & N/A & N/A &N/A \\
PCA NN     & 96.05\%  & 96.21\%   & 96.05\% & 95.92\%   
\end{tabular}
\caption{Performance evaluation of unsupervised \textit{(upper part)} and supervised \textit{(lower part)} models for contaminated time series localization step. Results for DBSCAN, sig-IF and SVM are not provided due to high computational cost.}
\label{tab:perfmethodsLocalizationReal}
\end{table}

\section{Conclusion}\label{section: conclusion}

We propose a two step approach for detecting anomalies on a panel of time series that can reflect a
wide variety of market risk factors. The first step aims at identifying the contaminated
time series, i.e. time series with anomalies. The second step focuses on the localization of the
anomaly among the observed values of the identified contaminated time series.
As preprocessing, our methodology integrates the extraction of features from the time series with
PCA. This part of the method proves to be essential, as it provides the models
with inputs on which the distinction between abnormal/contaminated and normal instances is
eased, while also ensuring the stationarity of the model (time series) inputs. Another key point of
the approach is the calibration of the cut-off value, the key parameter in the identification of
contaminated time series, by means of a feedforward neural network with a customized loss function.
The proposed approach suggests an imputation value, however this value is strongly influenced
by the abnormal value. Therefore, basic imputation approaches with similar complexity are preferred.
Our numerical experiments show not only that our approach outperforms
baseline anomaly detection models, but also show the real benefit that could be gained from
applying it as a data cleaning step preliminary to VaR computations.
Future research could focus on the replacement of PCA by partial least
squares (PLS) or deep PLS \citep{polson2021deep} for endogeniging the features extraction stage. Regarding downstream tasks, our approach might be of special interest for reverse stress tests \citep{BelliniEichhornMayenberger20book}.

\appendix

\section{Literature Review}\label{literatureReview}

We start with a  review of the anomaly detection literature. See also Section \ref{apx:AppendixA} for a more technical presentation of some of the below-mentioned algorithms.

\subsection{Baseline Algorithms}

Anomaly detection aims at finding an \textit{``observation that deviates so much from other observations as to arouse suspicion that it was generated by a different mechanism"} \citep{hawkins1980identification}. The baseline anomaly detection algorithms, described in \citep{chandola2009anomaly}, struggle to identify anomalies in time series, mainly because their assumptions are invalidated. If we consider models built for spatial data, a major assumption of these models is that observations are independent, whereas, for time series, high dependency exists between different time stamps. Clustering-based approaches, like the density based spatial clustering with noise (DBSCAN) method, are particularly impacted by this aspect: if an anomaly occurs at a given time stamp and is followed by incorrect
values, clustering-based approaches consider that the observations of the time series belong to two different clusters and thus fail to identify the anomaly. Another limitation when considering this type of techniques is the choice of the similarity metric used for data clustering. This task, although being a crucial pillar of these approaches, is not trivial and becomes very challenging for high dimensional problems. 



\subsection{Statistical Approach to Anomaly Detection}
The statistical techniques for anomaly detection can be split in two families: statistical tests and predictive models. Both suffer from the curse of dimensionality and model/data mismatch. When anomaly detection relies on hypothesis tests, it usually tests whether the observations are drawn from a known distribution \citep{zhang2017statistical}, supposing that the user knows the probability distribution of the normal observations. Such a parametric framework narrows down the scope of applicability of hypothesis tests, as the data does not always coincide with the assumed distribution. Moreover, the tests provided in the literature are not suitable in multivariate settings \citep{kurt2020real}. 
The statistical techniques relying on fitting a predictive model to each time series also require strong assumptions on the data. Predictive models are usually autoregressive (AR), moving average (MA), or ARMA models. Anomalies are then detected relatively to the forecasts suggested by the model \citep{chandola2009anomaly}. In these parametric approaches, some parameters have to be  specified again, starting with the order of the models. Selecting the optimal model parameters with respect to an information criterion is not always possible. Additionally, these models assume that the time series are homogeneous, i.e. drawn from the same distribution  \citep{laptev2015generic}. This is not always satisfied in the financial risk management case where several types of market risk factors are treated simultaneously.

\subsection{Score Based Anomaly Detection Models}
Additional anomaly detection challenges are of general concern.  Most of anomaly detection algorithms are score-based, in the sense that these approaches return an anomaly score reflecting to which extent the observation is considered to be abnormal by the model. In order to decide whether an observation is abnormal or not, a cut-off value of the score has to be selected. Empirical approaches are often used, consisting in setting the cut-off value as a quantile or elbow point of the distribution of the anomaly score. However, the selected cut-off value according to such methods remains arbitrary. \citep{gaoconverting} propose to rely on a cost of misclassification based on a weighted classification accuracy. Approaches to calculate ``optimal'' weights are described in \citep{lu2019learning}, but they involve a heuristic grid search technique. Another alternative is to determine the cut-off value by cross validation on the training data \citep{saha2009snake}. Finally, some methods do not select any cut-off value, but are based instead on a contamination rate. However, fixing a cut-off value or deciding on a contamination rate is not so different.  

\subsection{Scarcity of Anomalies and Data Augmentation}

The scarcity of anomalies within the data sets is another typical problem in anomaly detection. Anomalies are, by definition, rare events, therefore they are under-represented in the data set used to fit the models. This under-representation is not helping in the design of a reliable model able to identify anomalies. Classical methods to overcome this issue consider data augmentation. These techniques aim at producing new synthetic samples that will ultimately enhance model performance, leveraging on a better representation of the features space. As reported in \citep{wen2020time}, time series can be augmented by using a simple transformation in time domain \citep{cui2016multi}, frequency domain \citep{gao2020robusttad}, or more advanced generative approaches involving deep learning techniques, such as recurrent generative adversarial networks \citep{esteban2017real}. In practice, the use of generative models for data augmentation of time series with anomalies presents two limitations. First, training such models requires a large number of samples to guarantee a satisfactory performance. While restricted Boltzman machines do not exhibit this problem, \citep{kondratyev2020data} have shown that they fail into fitting multivariate complex distributions with nonlinear dependence structure. A more fundamental limitation affects the very idea behind generative models. Such networks are trained to learn a given distribution. However, by definition, anomalies are different from each other and therefore there is not a distribution that characterises them.

\subsection{Supervised vs. Unsupervised learning }

Since anomalies are the realisations of atypical events for which the distribution is unknown, it seems quite natural to use unsupervised algorithms.  However, these approaches are deemed more suitable for learning complex patterns and are task specific. Moreover, \citep{gornitz2013toward} pointed out that they often do not present a high prediction performance, in particular in high dimensional settings \citep{ruff2019deep}. Indeed, the performance of unsupervised shallow anomaly detection algorithms depends upon a features engineering step. \citep{akyildirim2022applications} use signatures to extract features which then feed algorithms such as isolation forest. This combination of techniques is shown to over-perform benchmark approaches. However, the designed model is task specific (detection of pump and dumps attacks) and the features extraction step is only efficient when at least one explanatory variable is considered in the analysis. 
Moreover, the unavailability of labelled data makes the model building and its evaluation even more complex. As for supervised methods, the only limitation on which the literature tends to agree is their incapacity to generalize the learned patterns to new samples, which is the consequence of misrepresentation of anomalies among the training samples \citep{zhao2018xgbod}. However, the scarcity of labelled data can be sidestepped through the use of data augmentation techniques on the fraction of available labelled data. For these reasons the supervised learning framework is to be preferred even when only a small set of labelled data is available.  

\subsection{Anomaly Detection on Time Series}

Usually, anomaly detection models on time series have two main components. The first component aims at extracting a parsimonious yet expressive representation of the time series. Several approaches are suggested in the literature to deal with such features extraction. Recently, deep neural networks have been shown to suffer from overparametrization and to be often computationally expensive \citep{dempster2020rocket, akyildirim2022applications}. Path signatures are also computationally demanding. This could perhaps be alleviated by the random signatures \citep{compagnoni2022randomized}. However, the information extracted with signatures is of most interest when the considered paths are characterized by several variables. The resulting representation is then transformed into a (typically continuous) anomaly score which is in turn converted into a binary label \citep{braei2020anomaly}. 

\subsection{PCA and Anomaly Detection}
In the literature, PCA is usually used in anomaly detection for its dimension reduction properties. Anomalies are identified on the latent space, either by applying some anomaly detection algorithm or by assuming a given distribution on the principal component and identifying the anomalies relatively to a quantile \citep{shyu2006principal}. Assuming that a normal subspace representation of the data set can be constructed with the first $k$-components \citep{ringberg2007sensitivity}, anomaly detection is achieved by looking at the observations that cannot be expressed in terms of the first $k$-components \citep{bin2016abnormal}. Note that the particular power-fullness of auto-encoder, a non-linear PCA, as data compressor is not desirable herein, since auto-encoders compress all patterns including abnormal ones. While \citep{ding2016pca} claimed that the PCA-based models are stable with respect to their parameters, such as the number of principal components $k$ spanning the subspace or the cut-off level, \citep{ringberg2007sensitivity} found instead that PCA-based anomaly detection is sensitive to these parameters and to the amplitude of the anomalies. Indeed, the latter may undermine the construction of the normal subspace representation, in turn leading to misidentification of anomalies. In light of that, we take an extra care regarding these aspects, and appropriate tests are conducted to show that the proposed approach is not subject to these issues. In the end, with a relatively low number of features given by PCA, we are able to accurately describe the dynamics of market risk factors represented by times series. The reason standing behind that is the high correlation structure displayed by market risk factors. 


\section{State of the Art Anomaly Detection Models}\label{apx:AppendixA}

There are mainly two categories of machine learning anomaly detection models :  density-based models, where a distribution is used to fit the data and anomalies are defined relatively to this distribution, and depth-based models, which, instead of modeling the normal behaviour, isolate anomalies.

\subsection{Density-Based Models}

\paragraph{Density based spatial clustering with noise (DBSCAN)}
DBSCAN \citep{ester1996density,schubert2017dbscan} is an unsupervised clustering methodology that groups together comparable observations based on a similarity metric. Clusters are high density regions and are defined by the $\varepsilon$-neighbourhood of observations and by \textit{MinPts},  the minimum number of points required to be in a radius of $\varepsilon$ from an observation to form a dense region. An anomaly is any observation which has not $MinPts$ in its $\varepsilon$-neighbourhood nor appears in the $\varepsilon$-neighbourhood of other observations.

\paragraph{K-nearest neighbours (KNN)}
Usually used for classification purposes, KNN \citep{hand2007principles} is a supervised algorithm that can be used for anomaly detection as well. For each observation it selects its closest $K$ observations generally in terms of distance, but other similarity metrics can be considered.  The anomaly score of an observation is computed as function of its distances to $K$-nearest neighbours, weighted average of the distances for instance. The points with the highest anomaly scores are considered as anomalies. 

\paragraph{Support Vector Machines (SVM)} The ultimate aim of SVM \citep{cortes1995support} is to define a hyperplan that separates the data. The specificity of this hyperplan is that it maximizes the distances to the set of features representing each class. When the data is not linearly separable, a map $\phi$ is applied to the initial features vector so the data became linearly separable in the new space in which it was projected. 

\subsection{Depth-Based Models}

\paragraph{Isolation Forest (IF)}
The IF method \citep{hariri2019extended} applies a depth approach to detect anomalies. The algorithm is based on the idea that anomalies are easier to isolate and thus will be isolated closer to the root, while normal observations are isolated much further from the root. The algorithm uses random decision trees to separate the observations. The anomaly score is calculated as the path length to isolate the observation. As it is defined in the algorithm, the anomaly score will be closer to 1 for anomalies and $\ll 1$ for normal observations. This allows ranking the observations, from the most abnormal observation to the most regular one. However, the choice of an explicit cut-off between this two type of instances is not obvious.  

\medskip

\paragraph{Local Outlier Factor (LOF)}
The LOF algorithm  \citep{breunig2000lof,alghushairy2020review} tries to assess the isolation of one observation relatively to the rest of the data set, before flagging it as an anomaly. This model relies on the concept of local density. The anomaly score of an observation $x$ is its local outlier factor, which quantifies how dense is the location area of $x$ compared to the one of its neighbours. Hence, for each observation, $\mathrm{LOF}_k(x) \approx 1$ means that the density of observations around $x$ is similar to the one of its neighbours, therefore $x$ could not be considered as an isolated observation. $\mathrm{LOF}_k(x) \gg 1$, instead, shows that the density of $x$ is lower compared to its neighbours, hence $x$ should be flagged as anomaly

\section{The Data Stationary Issue}\label{apx:AppendixB}
Assume that a process $\left(\varepsilon^{i}_t\right)$ satisfies the following representation 

\begin{align*}
    \Delta \varepsilon^{i}_t = \gamma \varepsilon^{i}_{t-1} + \theta_1 \Delta \varepsilon^{i}_{t-1}+\dots{}+\theta_{p-1} \Delta \varepsilon^{i}_{t-p+1} + z_t,
\end{align*}
where $p$ is the lag order, $\Delta$ is the difference operator, i.e $\Delta \varepsilon_t = \varepsilon_t - \varepsilon_{t-1} $, 
and $z_t$ is a white noise. The stationarity of $(\varepsilon^{i}_t)$ is shown using the augmented Dickey-Fuller test \citep{fuller2009introduction}. Namely, the process $\varepsilon^{i}$ is  stationary if there is a unit root, i.e. $\gamma=0$. Therefore, for each $\varepsilon^{i}$ the test is carried under the null hypothesis $
    \mathcal{H}_0 :  \gamma =0 \; \text{against} \;\mathcal{H}_1:\gamma<0$.

\begin{table}[H]
\centering
\begin{tabular}{c|c|c|c|c|c|c|c}

Set   & Mean & Standard dev. & Min & 25\% &50\% &75\% &Max \\ \hline
Train & 1.26e-14  & 1.74e-13 & 6.63e-24 & 3.46e-20  & 1.92e-18  & 3.74e-17  & 3.88e-12 \\ \hline
Test  & 1.51e-14  & 4.83-13 & 6.20e-30 & 4.23e-19  & 8.45e-18  & 1.65e-16  & 4.35e-11 \\ 
\end{tabular}
\caption{Descriptive statistics on p-values for train and test set time series $\bm{\varepsilon}$.}
\end{table}
If the p-values are lower than the significance level, then the null hypothesis is rejected for all the reconstruction errors and we conclude that the time series we work with do not suffer from the non-stationarity issue.

\medskip

\nocite{*}
\bibliographystyle{chicago}
\bibliography{mybib}

\end{document}